\documentclass[useAMS,usenatbib]{mn2e}
\usepackage{graphicx}
\title[Evolution of gas in Fornax]{Estimating the evolution of gas in the Fornax 
dwarf spheroidal galaxy from its star formation history: an illustrative example}
\author[Z. Yuan, Y.-Z. Qian and Y. P. Jing]{Zhen Yuan$^{1}$\thanks{E-mail:
yuan@physics.umn.edu (ZY); qian@physics.umn.edu (YZQ); ypjing@sjtu.edu.cn (YPJ)}, 
Y.-Z. Qian$^{1}$\footnotemark[1]\thanks{Visiting Chair Professor, Center for Nuclear Astrophysics, 
Department of Physics and Astronomy, Shanghai Jiao Tong University, Shanghai 200240, China }
and Y. P. Jing$^{2}$\footnotemark[1]\\
$^{1}$School of Physics and Astronomy, University of Minnesota, Minneapolis, MN 55455, USA\\
$^{2}$IFSA Collaborative Innovation Center, Center for Astronomy and Astrophysics, 
Department of Physics and Astronomy,\\ 
Shanghai Jiao Tong University, Shanghai 200240, China}

\begin{document}


\pagerange{\pageref{firstpage}--\pageref{lastpage}} \pubyear{2015}

\maketitle

\label{firstpage}

\begin{abstract}
We propose that detailed data on the star formation history of a dwarf spheroidal galaxy 
(dSph) may be used to estimate the evolution of the total mass $M_g(t)$ for cold gas
in its star-forming disk. Using Fornax as an illustrative example, we estimate its $M_g(t)$
and the corresponding net gas flow rate $\Delta F(t)$ assuming a global star formation 
rate $\psi(t)=\lambda_*(t)[M_g(t)/M_\odot]^\alpha$ consistent with observations of nearby 
galaxies. We identify the onset of the transition in $\Delta F(t)$ from a net inflow to a net 
outflow as the time $t_{\rm sat}$ at which the Fornax halo became a Milky Way 
satellite and estimate the evolution of its total mass $M_h(t)$ at $t< t_{\rm sat}$ using 
the median halo growth history in the $\Lambda$CDM cosmology and its present mass 
within the half-light radius derived from observations. We examine three different cases 
of $\alpha=1$, 1.5, and 2, and justify the corresponding $\lambda_*(t)$ by comparing 
the gas mass fraction $f_g(t)=M_g(t)/M_h(t)$ at $t< t_{\rm sat}$ with results from 
simulations of gas accretion by halos in a reionized universe. We find that the Fornax 
halo grew to $M_h(t_{\rm sat})\sim 2\times 10^9\,M_\odot$ at $t_{\rm sat}\sim 5$ or 
8~Gyr, in broad agreement with previous studies using data on its stellar kinematics
and its orbital motion. We describe qualitatively the evolution of Fornax as a 
satellite and discuss potential extension of our approach to other dSphs.
\end{abstract}


\section{Introduction}
As a result of hierarchical structure formation, the cold dark matter 
(CDM) halo associated with a main galaxy has many subhalos. In principle, this 
can account for the dwarf spheroidal galaxies (dSphs) orbiting
the Milky Way (MW). However, due to the complicated 
gas dynamics involved in star formation and the resulting feedback, 
it is far from straightforward to associate the subhalos found in CDM simulations 
with the actual dSphs possessing stellar populations. A prominent example is 
the so-called too-big-to-fail (TBTF) problem: the largest subhalos of an MW-like 
halo appear unsuitable for hosting the brightest MW dSphs because of
incompatible mass distributions (e.g., \citealt*{boylan}).
Proposed solutions to this problem mostly call for either modifying 
the nature of DM (e.g., \citealt{herpich}) or 
including baryons and the associated gas dynamics in CDM simulations
(see \citealt{pontzen} for a review and \citealt{sawala} for recent simulation results
on the Local Group galaxies). It was also shown that there might not be
a TBTF problem if the CDM halo of the MW has a relatively low mass
of $\approx 10^{12}\,M_\odot$ (e.g., \citealt{wang}). In any case, gas dynamics
is the crucial bridge linking DM halos with observed galaxies. 
 
In this paper we propose an empirical approach to study halo evolution and 
global gas dynamics of dSphs. We use Fornax, the brightest MW dSph, as an 
illustrative example. 
Our approach is in the same spirit as that of \cite{qw}, who used the data on
metallicity distributions of MW dSphs to infer the generic evolution of their gas.
We show that the global star formation rate (SFR) $\psi(t)$ of Fornax derived 
from observations by \cite{deboer} may be used to estimate its total gas mass 
$M_g(t)$ and the corresponding net gas flow rate $\Delta F(t)$ as functions of 
time $t$. We consider two regimes of gas evolution separated by the time 
$t_{\rm sat}$, at which Fornax ceased evolving independently and became an 
MW satellite. We assume $\psi(t)=\lambda_*(t)[M_g(t)/M_\odot]^\alpha$ with
$\lambda_*(t)=\lambda_*(t_{\rm sat})$ for $t>t_{\rm sat}$.
We consider that the accretion of Fornax by the MW caused the former's global 
gas flow to change rapidly from a net inflow to a net outflow and identify the 
onset of this transition in $\Delta F(t)$ as $t_{\rm sat}$. We assume that the total 
mass $M_h(t)$ of the Fornax halo at $t< t_{\rm sat}$ follows the median halo 
growth history in the $\Lambda$CDM cosmology and determine $M_h(t_{\rm sat})$ 
by requiring that the corresponding density profile give the mass enclosed within 
the half-light radius as derived from observations.

Considering the central role of the assumed global star formation law (SFL)
$\psi(t)=\lambda_*(t)[M_g(t)/M_\odot]^\alpha$ in our approach, we examine three different 
cases of $\alpha=1$, 1.5, and 2, and justify the corresponding $\lambda_*(t)$
by comparing the gas mass fraction $f_g(t)=M_g(t)/M_h(t)$ at $t< t_{\rm sat}$ with the 
results from cosmological simulations
of gas accretion by halos in a reionized universe. For each case we also perform 
statistical realizations of the data on $\psi(t)$ to gauge how uncertainties in $\psi(t)$ affect
our results, especially $\Delta F(t)$. Using representative results for the
baseline case of $\alpha=1.5$, we give a qualitative description of Fornax's overall gas
evolution, especially its gas loss through a combination of ram-pressure stripping and
tidal interaction with the MW at $t>t_{\rm sat}$.

This paper is organized as follows. In \S\ref{sec-sf} we discuss theoretical and 
empirical SFLs and motivate our assumed SFL for dSphs. In \S\ref{sec-gas} we use 
Fornax as an illustrative example to estimate $M_g(t)$, $\Delta F(t)$, $t_{\rm sat}$, 
and $M_h(t_{\rm sat})$ from its star formation history (SFH), 
approximately taking into account uncertainties in the SFH. We also
compare our estimates of $t_{\rm sat}$ and $M_h(t_{\rm sat})$ with relevant
results from previous studies, justify choices of $\lambda_*(t)$ by considering 
gas evolution at $t< t_{\rm sat}$, and examine how our 
results are affected by our assumed SFLs. In \S\ref{sec-sat} we 
qualitatively relate $\Delta F(t)$ at $t>t_{\rm sat}$ to the orbital motion of Fornax 
as an MW satellite, and discuss how it lost all of its gas through a combination of 
ram-pressure stripping and tidal interaction with the MW. We summarize our results 
and discuss potential extension of our approach to other dSphs in \S\ref{sec-dc}.

\section{SFLs and Star Formation in a dSph}
\label{sec-sf}
An in-depth study of the halo evolution and the associated gas dynamics 
of a dSph would require simulations that incorporate hierarchical structure formation 
in the $\Lambda$CDM cosmology and treat a wide range of physical processes, such
as gas accretion by the hosting halo before and after reionization of the universe, 
cooling and condensation of gas to form stars, conversion of cold into hot gas by 
radiation and supernova (SN) explosion, and gas expulsion from the halo
(see e.g., \citealt{somer} for a review).
A representative example of this approach is the recent work of \cite{shen}, who 
simulated the formation and evolution of seven field dwarf galaxies with present-day 
virial masses of $M_{\rm vir}=4.4\times 10^8$--$3.6\times 10^{10}\,M_\odot$.
Our approach here is empirical and phenomenological, especially regarding star
formation and gas inflows and outflows. Central to our approach is the SFL
relating the SFR to gas properties. Therefore, we start with a brief discussion of 
theoretical and empirical SFLs 
(see e.g., \citealt{kenn12,krum14} for comprehensive reviews).

\subsection{Theoretical and empirical SFLs}
Star formation is most directly associated with molecular gas (see e.g.,
\citealt{gao} and \citealt{kth07} for discussion of observations using different
molecular tracers). For a cloud with
a total mass $M_{\rm mol}$ of molecular gas (predominantly consisting of 
H$_2$ molecules and the associated He atoms),
a theoretical estimate of the SFR (e.g., \citealt{kt07}) is
\begin{equation}
\frac{dM_*}{dt}=\epsilon_{\rm ff}\frac{M_{\rm mol}}{t_{\rm ff}},
\label{eq-sflm}
\end{equation}
where $M_*$ is the mass of stars formed,
\begin{equation}
t_{\rm ff}\equiv\sqrt{\frac{3\pi}{32G\rho_{\rm mol}}}
\label{eq-tff}
\end{equation}
is the free-fall time at the density $\rho_{\rm mol}$ of the molecular gas,
$G$ is the gravitational constant, and $\epsilon_{\rm ff}$ is the fraction of the 
molecular gas turned into stars during one free-fall time. \cite{kt07} obtained 
$\epsilon_{\rm ff}\sim 10^{-2}$ from observations covering a wide range of 
$n_{\rm mol}\equiv\rho_{\rm mol}/m_p\sim 10$--$10^4$~cm$^{-3}$, 
where $m_p$ is the proton mass. An alternative form of Eq.~(\ref{eq-sflm}) is
\begin{equation}
\Sigma_{\rm SFR}=\epsilon_{\rm ff}\frac{f_{{\rm H}_2}\Sigma_g}{t_{\rm ff}},
\label{eq-sfl}
\end{equation}
where $\Sigma_{\rm SFR}$ is the SFR per unit area, $\Sigma_g$ is the
surface mass density of gas in all forms, and $f_{{\rm H}_2}$ is the fraction 
of H mass in H$_2$ molecules. As star formation typically occurs in gaseous 
disks, Eq.~(\ref{eq-sfl}) is more convenient to use, especially for comparison 
with observations. Both Eqs.~(\ref{eq-sflm}) and (\ref{eq-sfl}) can be applied to 
individual star-forming regions and in the disk-averaged form, to individual 
galaxies (see review by \citealt{krum14}). For illustration, we compare them 
with two observational results below. 

From observations of nearby galaxies on sub-kpc scales, 
\cite{bigiel} obtained an SFL that applies to local regions within a galaxy:
\begin{equation}
\frac{\Sigma_{\rm SFR}}{M_\odot\ {\rm yr}^{-1}\ {\rm kpc}^{-2}}=10^{-2.1\pm 0.2}
\left(\frac{\Sigma_{{\rm H}_2}}{10\,M_\odot\ {\rm pc}^{-2}}\right)^{1.0\pm 0.2},
\label{eq-sflb}
\end{equation}
where $\Sigma_{{\rm H}_2}$ is the surface mass density of H$_2$ molecules 
in a star-forming region. The above SFL covers
$\Sigma_{{\rm H}_2}\sim 3$--$50\,M_\odot$~pc$^{-2}$.
Taking the central values of the fitting parameters, we can rewrite it as
$dM_*/dt\sim M_{\rm mol}/t_{\rm depl}\sim 1.4M_{{\rm H}_2}/t_{\rm depl}$, 
where $M_{{\rm H}_2}$ is the total mass of H$_2$ molecules and
$t_{\rm depl}\sim 2$~Gyr is the time needed for star formation to deplete 
all the molecular gas. This depletion time can be understood from
$t_{\rm depl}\sim t_{\rm ff}/\epsilon_{\rm ff}$ [see Eq.~(\ref{eq-sflm})]. 
The star-forming regions studied by \cite{bigiel} were resolved to 750~pc. 
Assuming a reasonable scale-height ${\cal H}\sim 100$~pc for such regions
with a typical $\Sigma_{{\rm H}_2}\sim 10\,M_\odot$~pc$^{-2}$, we obtain 
$\rho_{\rm mol}\sim 1.4\Sigma_{{\rm H}_2}/{\cal H}\sim 0.14\,M_\odot$~pc$^{-3}$
($n_{\rm mol}\sim 5.7$~cm$^{-3}$), which corresponds to
$t_{\rm ff}\sim 22$~Myr [see Eq.~(\ref{eq-tff})] or 
$t_{\rm depl}\sim 2.2$~Gyr for $\epsilon_{\rm ff}\sim 10^{-2}$.

Based on observations of 61 normal spiral galaxies and 36 starburst galaxies,
\cite{kenn} found a tight correlation between $\Sigma_{\rm SFR}$ and $\Sigma_g$, 
both of which are averaged over the disk of an individual galaxy. This global SFL is
\begin{equation}
\frac{\Sigma_{\rm SFR}}{M_\odot\ {\rm yr}^{-1}\ {\rm kpc}^{-2}}=
(2.5\pm0.7)\times 10^{-4}
\left(\frac{\Sigma_g}{M_\odot\ {\rm pc}^{-2}}\right)^{1.4\pm 0.15},
\label{eq-sflk1}
\end{equation}
where $\Sigma_g$ ranges from $\sim 10$ to $10^4\,M_\odot$~pc$^{-2}$ 
for the galaxies used. The power-law form 
$\Sigma_{\rm SFR}\propto\Sigma_g^\alpha$ was first suggested by
\cite{schm}, and Eq.~(\ref{eq-sflk1}) is commonly referred to as
the Kennicutt-Schmidt law.
\cite{kenn} also found an alternative form of this SFL:
\begin{equation}
\frac{\Sigma_{\rm SFR}}{M_\odot\ {\rm yr}^{-1}\ {\rm kpc}^{-2}}=
1.7\times 10^{-2}\left(\frac{\Sigma_g}{M_\odot\ {\rm pc}^{-2}}\right)
\left(\frac{\rm yr}{\tau_{\rm dyn}}\right),
\label{eq-sflk2}
\end{equation}
where $\tau_{\rm dyn}$ is the dynamic timescale taken to be the orbit time 
at half of the outer radius of the
star-forming disk and ranges from $\sim 10^7$ to $\sim 10^8$~yr.
Detailed models to explain Eqs.~(\ref{eq-sflk1}) and
(\ref{eq-sflk2}) were given by \cite{km05}. Here we consider the limit of 
$\Sigma_g\ga 10^2\,M_\odot$~pc$^{-2}$, for which gas predominantly
consists of H$_2$ molecules \citep{km05} and Eq.~(\ref{eq-sfl}) 
can be applied to give
$\Sigma_{\rm SFR}\sim\epsilon_{\rm ff}\Sigma_g/t_{\rm ff}$.
Evaluating $t_{\rm ff}$ at the gas density $\rho_g\sim\Sigma_g/{\cal H}$, 
we obtain 
\begin{eqnarray}
\frac{\Sigma_{\rm SFR}}{M_\odot\ {\rm yr}^{-1}\ {\rm kpc}^{-2}}&\sim&
1.2\times 10^{-4}\left(\frac{\epsilon_{\rm ff}}{10^{-2}}\right)
\left(\frac{\rm 100\ pc}{\cal H}\right)^{0.5}\nonumber\\
&&\times\left(\frac{\Sigma_g}{M_\odot\ {\rm pc}^{-2}}\right)^{1.5},
\end{eqnarray}
which is in good agreement with Eq.~(\ref{eq-sflk1}) for 
$\epsilon_{\rm ff}\sim10^{-2}$ and disk scale-heights of 
${\cal H}\sim 100$~pc. A similar argument was used by \cite{kenn} 
to explain qualitatively the power-law form of Eq.~(\ref{eq-sflk1}).

\subsection{Examples of treating star formation in galaxy simulations}
One way to treat star formation in modeling galaxies is to use Eq.~(\ref{eq-sflm})
or its equivalent, Eq.~(\ref{eq-sfl}). In either case, estimates of $f_{{\rm H}_2}$ 
are required. For systems with sufficiently high SFRs, $f_{{\rm H}_2}$ depends on 
$\Sigma_g$, the metallicity, and the clumpiness of gas 
\citep*{kmt}. For systems with very low SFRs, $f_{{\rm H}_2}$ also depends on
$\Sigma_{\rm SFR}$ \citep{krum13}. Incorporating all the above dependences of
$f_{{\rm H}_2}$, \cite{krum13} showed that Eq.~(\ref{eq-sfl}) provides a good
description of the data on star formation for conditions ranging from poor to rich 
in molecular gas. 

However, many simulations of galaxy formation do not
evaluate $f_{{\rm H}_2}$ in implementing Eq.~(\ref{eq-sflm}) or (\ref{eq-sfl})
for star formation (see e.g., the review by \citealt{somer}). Instead, stars 
are assumed to form when gas has cooled below some temperature $T_{\rm max}$ 
and condensed to some threshold number density of atoms $n_{\rm min}$. 
For example, \cite{stinson} chose $T_{\rm max}=1.5\times 10^4$~K and 
$n_{\rm min}=0.1$~cm$^{-3}$. For a region with a total mass $M_g$ of the gas 
that satisfies the conditions and has a dynamic timescale 
$t_{\rm dyn}\propto(G\rho_g)^{-1/2}$, stars were assumed to form stochastically 
at a rate
\begin{equation}
\frac{dM_*}{dt}=0.05\frac{M_g}{t_{\rm dyn}}.
\label{eq-sflst}
\end{equation}
In comparing the above equation with Eq.~(\ref{eq-sflm}), 
$M_g$ is a proxy for $M_{\rm mol}$ 
while the numerical coefficient and $t_{\rm dyn}$ play the roles of 
$\epsilon_{\rm ff}$ and $t_{\rm ff}$, respectively. 
More recent simulations of \cite{shen} adopted $T_{\rm max}=10^4$~K, 
$n_{\rm min}=10^2$~cm$^{-3}$, and an SFR per unit volume
\begin{equation}
\frac{d\rho_*}{dt}=0.1\frac{\rho_g}{t_{\rm dyn}},
\end{equation}
where $\rho_*$ is the density of stars formed. The above SFL is 
qualitatively the same as Eq.~(\ref{eq-sflst}).

\subsection{Assumed dSph SFL based on gas mass}
\label{sec-sflfor}
As our goal is to gain some understanding of the overall picture for 
halo evolution and gas dynamics of dSphs in general and Fornax
in particular, we are interested in the relation of a dSph's global SFR
to its properties on the galactic scale. While we recognize the important 
role of molecular gas in star formation, our simple approach here will 
focus on the net cold gas in atomic
and molecular forms, which dominates the total gas mass 
in the star-forming disk. Based on observations of seven nearby spiral galaxies
resolved to 750~pc, \cite{bigiel} found that over the range of 
$\Sigma_{\rm H}\sim 1$--$10^2\,M_\odot$~pc$^{-2}$ for the net surface 
mass density of H atoms and H$_2$ molecules in star-forming regions,
$\Sigma_{\rm SFR}$ can be described by
\begin{equation}
\frac{\Sigma_{\rm SFR}}{M_\odot\ {\rm yr}^{-1}\ {\rm kpc}^{-2}}=10^{-2.39\pm 0.28}
\left(\frac{\Sigma_{\rm H}}{10\,M_\odot\ {\rm pc}^{-2}}\right)^{1.85\pm 0.70}.
\label{eq-sflbg}
\end{equation}
More recently, \cite{roy} carried out a spatially-resolved study of star-forming 
regions dominated by H atoms in nearby massive spiral and faint
dwarf irregular galaxies. They found a much tighter relation
\begin{equation}
\frac{\Sigma_{\rm SFR}}{M_\odot\ {\rm yr}^{-1}\ {\rm kpc}^{-2}}\sim
10^{-3}\left(\frac{\Sigma_g}{10\,M_\odot\ {\rm pc}^{-2}}\right)^{1.5}
\label{eq-sflroy}
\end{equation}
over the range of $\Sigma_g\sim 0.3$--$30\,M_\odot$~pc$^{-2}$.
In addition, the Kennicutt-Schmidt law in Eq.~(\ref{eq-sflk1}) describes the
global $\Sigma_{\rm SFR}$ for individual galaxies with
$\Sigma_g\sim 10$--$10^4\,M_\odot$~pc$^{-2}$. The above three empirical
results on SFLs have overlapping ranges of $\Sigma_g$ and are in 
broad agreement over these ranges within their uncertainties.
Guided by these results, we assume that the global SFR $\psi(t)$ 
in a dSph is related to the total mass $M_g(t)$ of gas in its star-forming disk as
\begin{equation}
\psi(t)=\frac{dM_*}{dt}=\lambda_*(t)\left[\frac{M_g(t)}{M_\odot}\right]^\alpha,
\label{eq-sfr}
\end{equation}
where $\lambda_*(t)$ is a rate function. 

We can use the effective area $A_{\rm disk}(t)$ of the star-forming disk to rewrite 
Eq.~(\ref{eq-sfr}) as
\begin{eqnarray}
\frac{\Sigma_{\rm SFR}(t)}{M_\odot\ {\rm yr}^{-1}\ {\rm kpc}^{-2}}&=&
\frac{\lambda_*(t)}{10^{1-8\alpha}\,M_\odot\ {\rm yr}^{-1}}
\left[\frac{A_{\rm disk}(t)}{10\ {\rm kpc}^2}\right]^{\alpha-1}\nonumber\\
&&\times\left[\frac{\Sigma_g(t)}{10\,M_\odot\ {\rm pc}^{-2}}\right]^\alpha.
\label{eq-ssfl}
\end{eqnarray}
For the above equation to agree with the empirical SFLs in Eqs.~(\ref{eq-sflk1}), 
(\ref{eq-sflbg}), and (\ref{eq-sflroy}), we take
\begin{equation}
\lambda_*(t)\sim 10^{-2-8\alpha}
\left[\frac{10\ {\rm kpc}^2}{A_{\rm disk}(t)}\right]^{\alpha-1}\,M_\odot\ {\rm yr}^{-1}
\label{eq-lam}
\end{equation}
and examine cases of $\alpha=1$, 1.5, and 2. We assume that $A_{\rm disk}(t)$
grew with time until a dSph became a satellite at $t=t_{\rm sat}$ and then 
remained fixed at $A_{\rm disk}(t_{\rm sat})$ during the subsequent SFH. 
As the majority of stars in the present-day Fornax are distributed within a region 
of $r_*\sim 2$~kpc in radius \citep{deboer}, we take 
$A_{\rm disc}(t_{\rm sat})\sim\pi r_*^2\sim 10$~kpc$^2$ and estimate
$\lambda_*(t_{\rm sat})\sim 10^{-2-8\alpha}\,M_\odot$~yr$^{-1}$. We will provide 
another justification for the choice of $\lambda_*(t_{\rm sat})$ in \S\ref{sec-gas}.
Note that the case of $\alpha=1$ is special because
the corresponding $\lambda_*(t)$ is independent of time [see Eq.~(\ref{eq-lam})]. 
In this case, $\lambda_*\sim 10^{-10}\,M_\odot$~yr$^{-1}$ 
is simply the fixed rate of gas consumption by star formation.

\section{Gas Mass, Net Gas Flow, and Halo Evolution for Fornax}
\label{sec-gas}
\begin{figure}
\includegraphics[width=84mm]{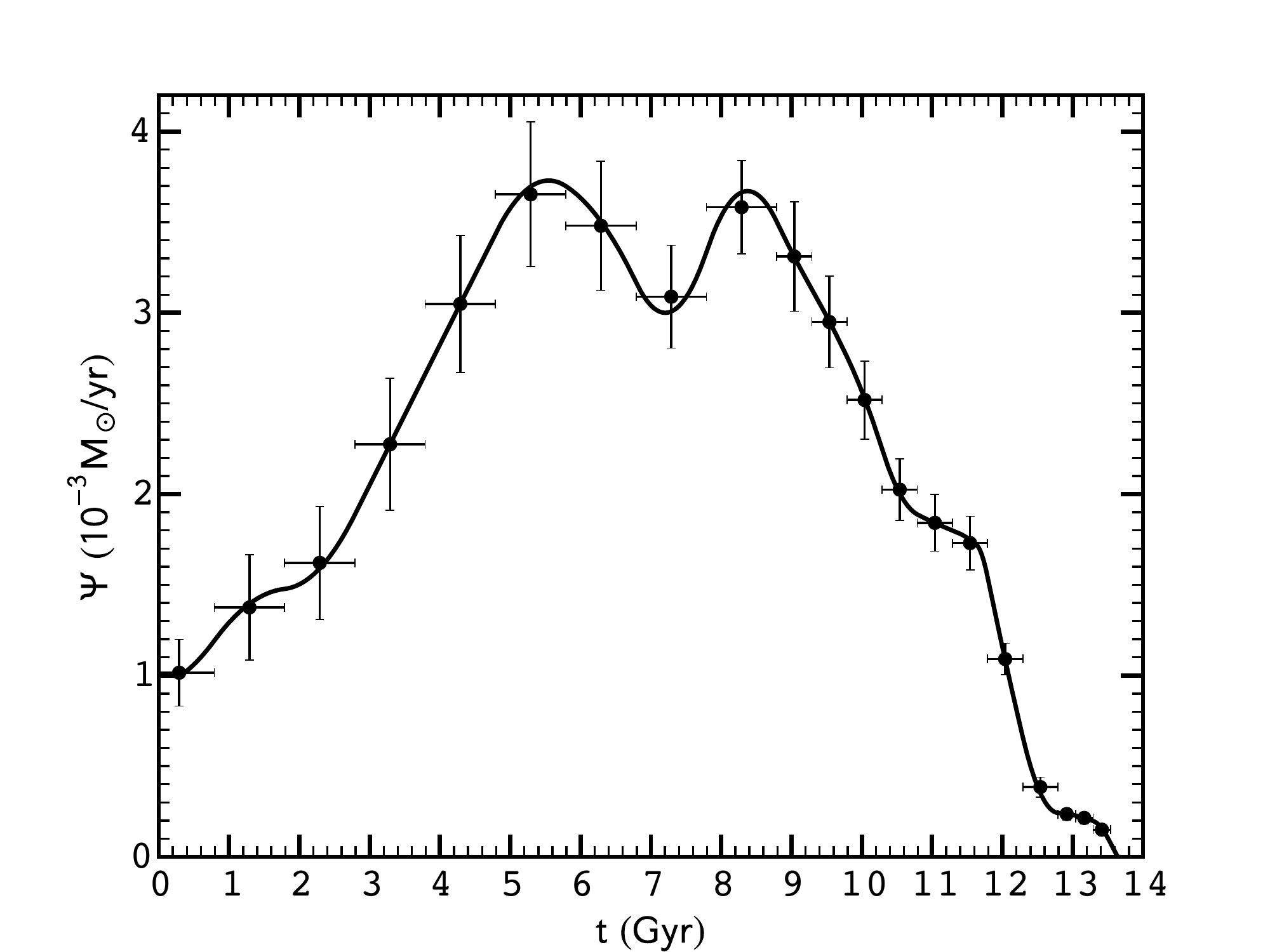}
\caption{Data (\protect\citealt{deboer}; filled circles with error bars) on Fornax's global 
star formation rate $\psi$ as a function of time $t$ since the big bang.
The data are binned according to time (converted from stellar age) and the 
horizontal error bar represents the size of each time bin. The filled circles
give the mean for the average star formation rate $\bar\psi$ in each bin and 
the vertical error bar represents the $1\sigma$ uncertainty in $\bar\psi$. 
A quadratic-spline fit (see Appendix~\ref{sec-fit}) corresponding to the filled
circles is shown as the solid curve. We take the centroid of the first time bin 
at $t\approx 0.29$~Gyr as the onset and $t\approx 13.6$~Gyr given by the 
fit as the end of star formation in Fornax.}
\label{fig-sfr}
\end{figure}

The data of \cite{deboer} on Fornax's $\psi(t)$
are binned across 13.75~Gyr and give the average SFR $\bar\psi$
for each time bin. As discussed in Appendix~\ref{sec-fit}, we obtain a smooth 
$\psi(t)$ using a quadratic-spline fit that conserves the total number of stars 
formed in each bin and guarantees the continuity of $\psi(t)$ and $d\psi/dt$. 
The fitted $\psi(t)$ corresponding to the mean for $\bar\psi$ in each bin is 
shown along with the data in Fig.~\ref{fig-sfr}. 

By choosing an appropriate 
$\lambda_*(t)$, we estimate $M_g(t)$ from Eq.~(\ref{eq-sfr}) as
\begin{equation}
\frac{M_g(t)}{M_\odot}=\left[\frac{\psi(t)}{\lambda_*(t)}\right]^{1/\alpha}.
\label{eq-mg}
\end{equation}
We first fix $\lambda_*(t)=\lambda_*(t_{\rm sat})$. As discussed in 
\S\ref{sec-sflfor}, comparison of our assumed SFL with 
empirical results gives an estimate of
$\lambda_*(t_{\rm sat})\sim 10^{-2-8\alpha}\,M_\odot$~yr$^{-1}$.
We then seek to identify $t_{\rm sat}$, the time at which Fornax
became an MW satellite, from the evolution of the net gas flow rate
\begin{equation}
\Delta F(t)\equiv\frac{dM_g}{dt}+\psi(t).
\label{eq-df}
\end{equation}
A net inflow onto or outflow from the star-forming disk of Fornax corresponds
to $\Delta F(t)>0$ or $\Delta F(t)<0$, respectively. We expect that the net gas 
inflow would be suppressed at $t\sim t_{\rm sat}$ due to gas loss through 
ram-pressure stripping and tidal interaction with the MW (\citealt{mayer}; 
see also \S\ref{sec-sat}).

Crucial to the identification of $t_{\rm sat}$ through features of $\Delta F(t)$ 
are the uncertainties in $\psi(t)$. Ideally, the permitted smooth $\psi(t)$ 
should be obtained from the raw observational data by taking into account
correlated errors in each time bin. Such an analysis is beyond our scope
here. Upon advice from T. J. L. de Boer, we give instead an approximate 
treatment as follows. We first make a Monte Carlo realization of the data 
by sampling $\bar\psi$ for each time bin according to a Gaussian 
distribution with the mean and standard deviation specified by the 
corresponding mean and $1\sigma$ uncertainty shown in Fig.~\ref{fig-sfr}. 
The realization is accepted only if it gives 
$(3.12\pm0.05)\times 10^7\,M_\odot$ 
for the total mass of stars formed (personal communications from 
T. J. L. de Boer). Out of a total of $10^4$ realizations, only $3806$ satisfy 
the constraint. For each accepted realization, we obtain a quadratic-spline 
fit to $\psi(t)$ and use it to calculate $M_g(t)$ and $\Delta F(t)$ 
from Eqs.~(\ref{eq-mg}) and (\ref{eq-df}), respectively.

\subsection{Baseline case}
\label{sec-base}
\begin{figure}
\includegraphics[width=84mm]{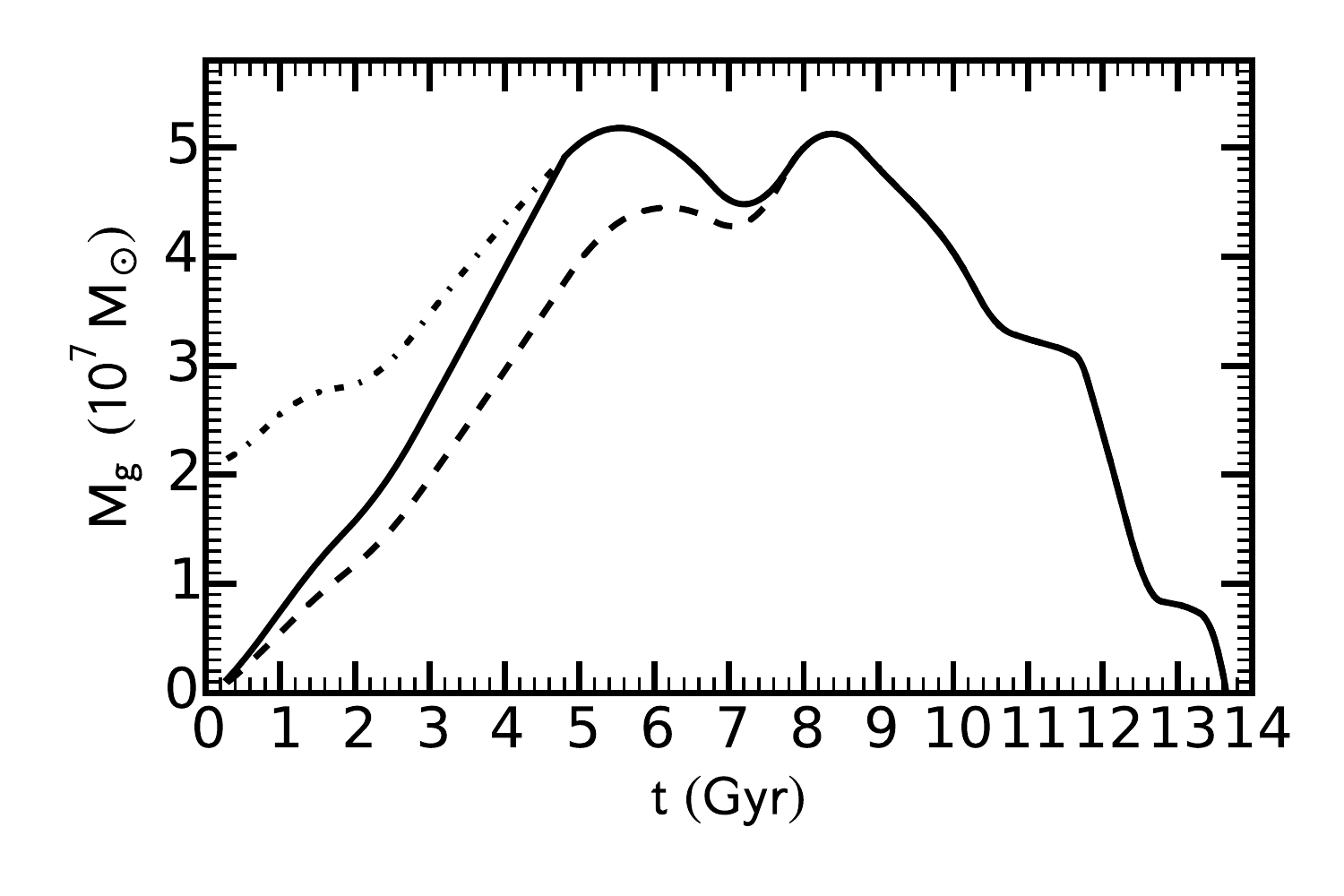}
\caption{Total mass $M_g$ for cold gas in Fornax's star-forming disk 
as a function of time $t$ derived from Eq.~(\ref{eq-mg}) and the smooth $\psi(t)$
shown in Fig.~\ref{fig-sfr} for the baseline case of $\alpha=1.5$ and
$\lambda_*(t_{\rm sat})=10^{-14}\,M_\odot$~yr$^{-1}$.
The top curve assumes a disk of fixed size with 
$\lambda_*(t)=\lambda_*(t_{\rm sat})$. The middle (bottom) curve is for a disk 
growing until $t_{\rm sat}\approx 4.8$ (7.8)~Gyr. The middle and top (bottom)
curves have the same $M_g(t)$ for $t>4.8$ (7.8)~Gyr.
Note that all curves reach $M_g=0$ at $t\approx 13.6$~Gyr,
the end of star formation in Fornax.}
\label{fig-mg}
\end{figure}

\begin{figure}
\includegraphics[width=84mm]{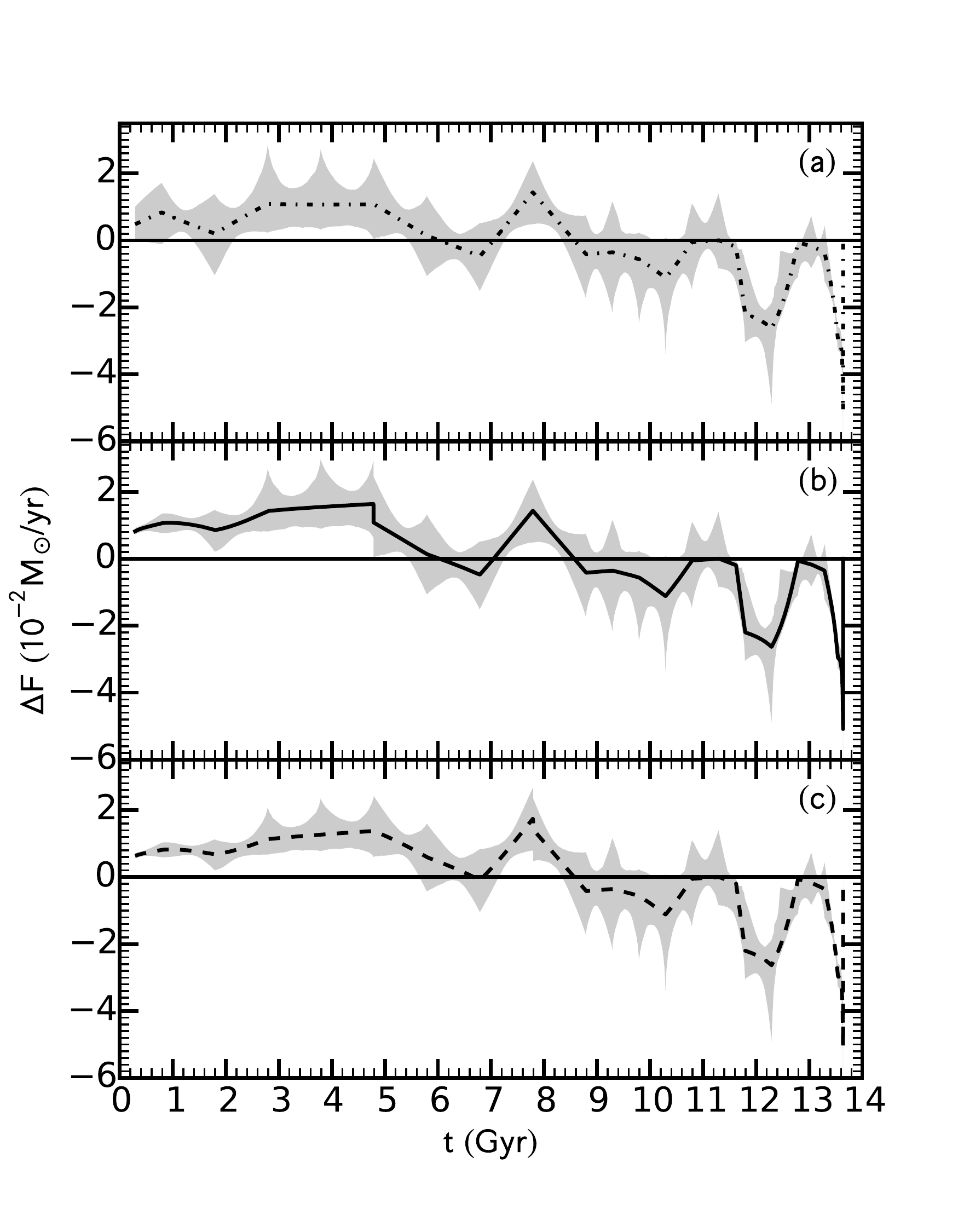}
\caption{Rate $\Delta F$ of net gas flow to or from Fornax's star-forming disk
as a function of time $t$ estimated from Eqs.~(\ref{eq-mg}) and (\ref{eq-df}) 
and the data shown in Fig.~\ref{fig-sfr} for the baseline case of
$\alpha=1.5$ and $\lambda_*(t_{\rm sat})=10^{-14}\,M_\odot$~yr$^{-1}$.
A net inflow or outflow corresponds to $\Delta F>0$ and $<0$, respectively. 
In each panel, the curve is calculated from the smooth $\psi(t)$ shown in
Fig.~\ref{fig-sfr} and the shaded region indicates the 68\% confidence interval.
Panel (a) assumes a disk of fixed size with 
$\lambda_*(t)=\lambda_*(t_{\rm sat})$. Panels (b) and (c) are for a disk 
growing until $t_{\rm sat}\approx 4.8$ and 7.8~Gyr, respectively. 
We attribute the onset of suppression of the net gas inflow at 
$t\approx 4.8$ or 7.8~Gyr to effects from 
Fornax's becoming an MW satellite. See text for details.}
\label{fig-df}
\end{figure}

We first discuss our baseline case of $\alpha=1.5$ and 
$\lambda_*(t_{\rm sat})=10^{-14}\,M_\odot$~yr$^{-1}$. 
The $M_g(t)$ calculated from the smooth $\psi(t)$ shown
in Fig.~\ref{fig-sfr} assuming that $\lambda_*(t)=\lambda_*(t_{\rm sat})$
is shown as the top curve in Fig.~\ref{fig-mg}.
The corresponding $\Delta F(t)$ is shown as the curve in 
Fig.~\ref{fig-df}a. The shaded region in this figure indicates the 68\% 
confidence interval for $\Delta F(t)$ estimated from the Monte Carlo
simulations constrained by the total mass of stars formed. 

We take $t_{\rm sat}$ to coincide with the onset of suppression of the net 
gas inflow. Based on the $\Delta F(t)$ shown in Fig.~\ref{fig-df}a, we consider 
two possibilities of $t_{\rm sat}\approx 4.8$ and 7.8~Gyr, respectively.
Using the Via Lactea II cosmological simulations, 
\cite{rocha} found that the infall times 
$t_{\rm infall}$ for the MW satellites can be constrained by their present-day 
kinematics (radial velocities, proper motions, and Galacto-centric positions). 
They found that $t_{\rm infall}\approx 4.8$--5.8 or 7.8--9.8~Gyr for Fornax, 
with the former range being more probable. It is intriguing that our estimates
of $t_{\rm sat}$ based on Fornax's SFH are consistent with theirs for 
$t_{\rm infall}$ based on Fornax's present-day kinematics.

For each choice of $t_{\rm sat}$,
we assume that the Fornax halo stopped growing at $t=t_{\rm sat}$ 
and that at $t<t_{\rm sat}$, its total mass $M_h(t)$ followed the median halo 
growth history in the $\Lambda$CDM cosmology. As discussed in 
Appendix~\ref{sec-mh}, we use the model of \cite{zhao} to estimate the 
dependence of the CDM density profile on the halo mass and collapse time,
and select the appropriate $M_h(t_{\rm sat})$ to obtain a mass
$M(<r_{1/2})=7.39^{+0.41}_{-0.36}\times 10^7\,M_\odot$ enclosed within 
the half-light radius $r_{1/2}=944\pm53$~pc as derived from observations 
\citep{wolf}. We find that $M_h(t_{\rm sat})\approx 1.8\times 10^9\,M_\odot$
and $2.4\times 10^9\,M_\odot$ for $t_{\rm sat}\approx 4.8$ and 7.8~Gyr, 
respectively. Using CDM mass distributions in subhalos from high-resolution
simulations of the Aquarius Project, \cite{strig} found a present-day halo 
mass of $7\times 10^8\,M_\odot$ for Fornax based on its stellar kinematics.
Considering the approximate nature of our approach and that the Fornax 
halo likely lost CDM at $t>t_{\rm sat}$ due to tidal interaction with the MW 
(see \S\ref{sec-sat}), we regard our estimates of $M_h(t_{\rm sat})$,
especially the lower one for $t_{\rm sat}\approx 4.8$~Gyr, as consistent 
with this more rigorous result on Fornax's present-day halo mass.

\begin{figure}
\includegraphics[width=84mm]{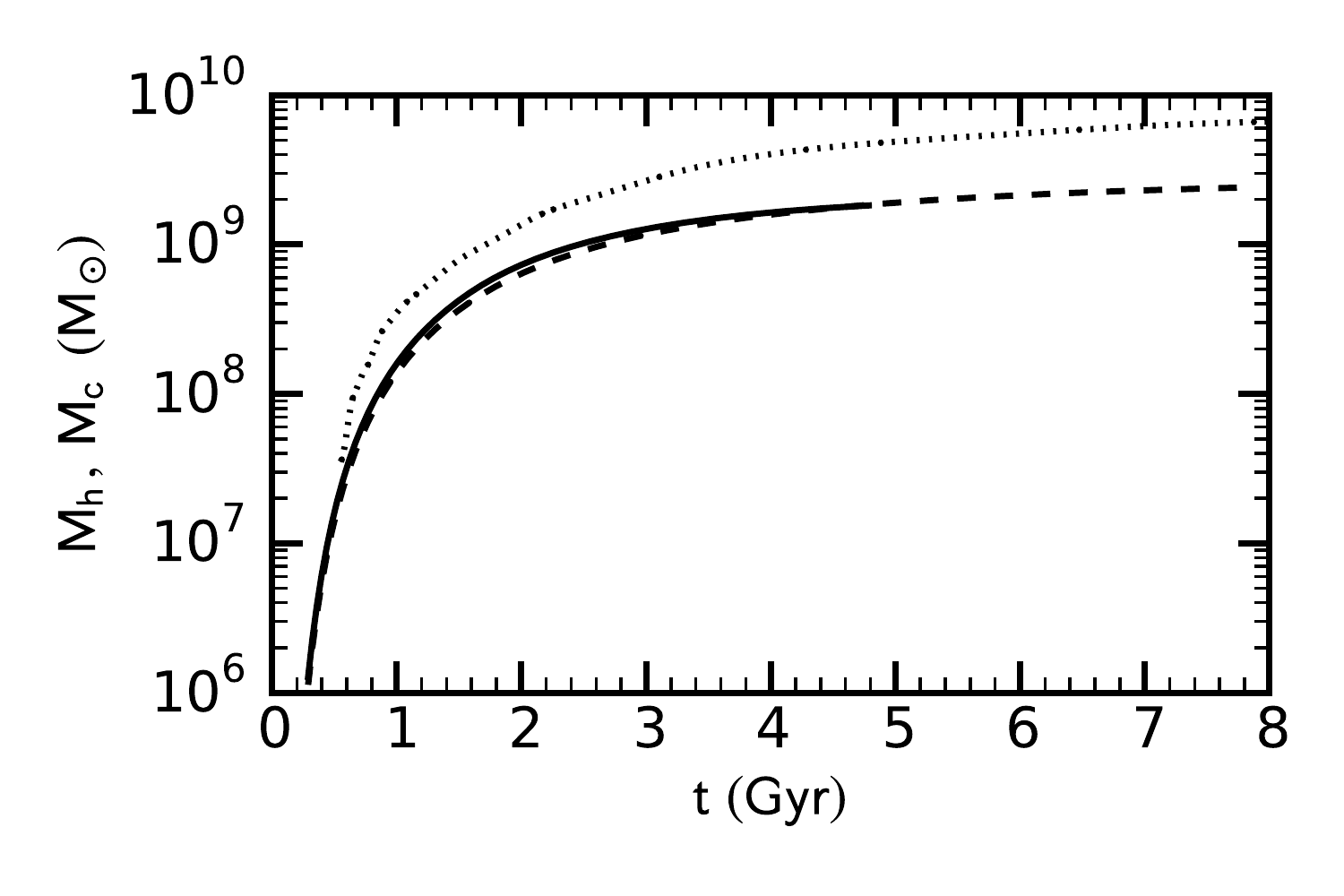}
\caption{Total mass $M_h$ of the Fornax halo as a function of time $t$ 
before it became an MW satellite at $t=t_{\rm sat}$. The solid (dashed) 
curve is for $t_{\rm sat}\approx 4.8$ (7.8)~Gyr. These results are 
estimated from the median halo growth history in the model of 
\protect\cite{zhao} and the present mass within the half-light radius as 
derived by \protect\cite{wolf} from observations 
(see Appendix~\ref{sec-mh}). The dotted curve shows the characteristic 
halo mass $M_c(t)$ used in Eq.~(\ref{eq-fb}), which gives the 
mean baryonic mass fraction for halos in a reionized universe based on 
simulations of \protect\cite{okamoto}.}
\label{fig-mh}
\end{figure}

\begin{figure}
\includegraphics[width=84mm]{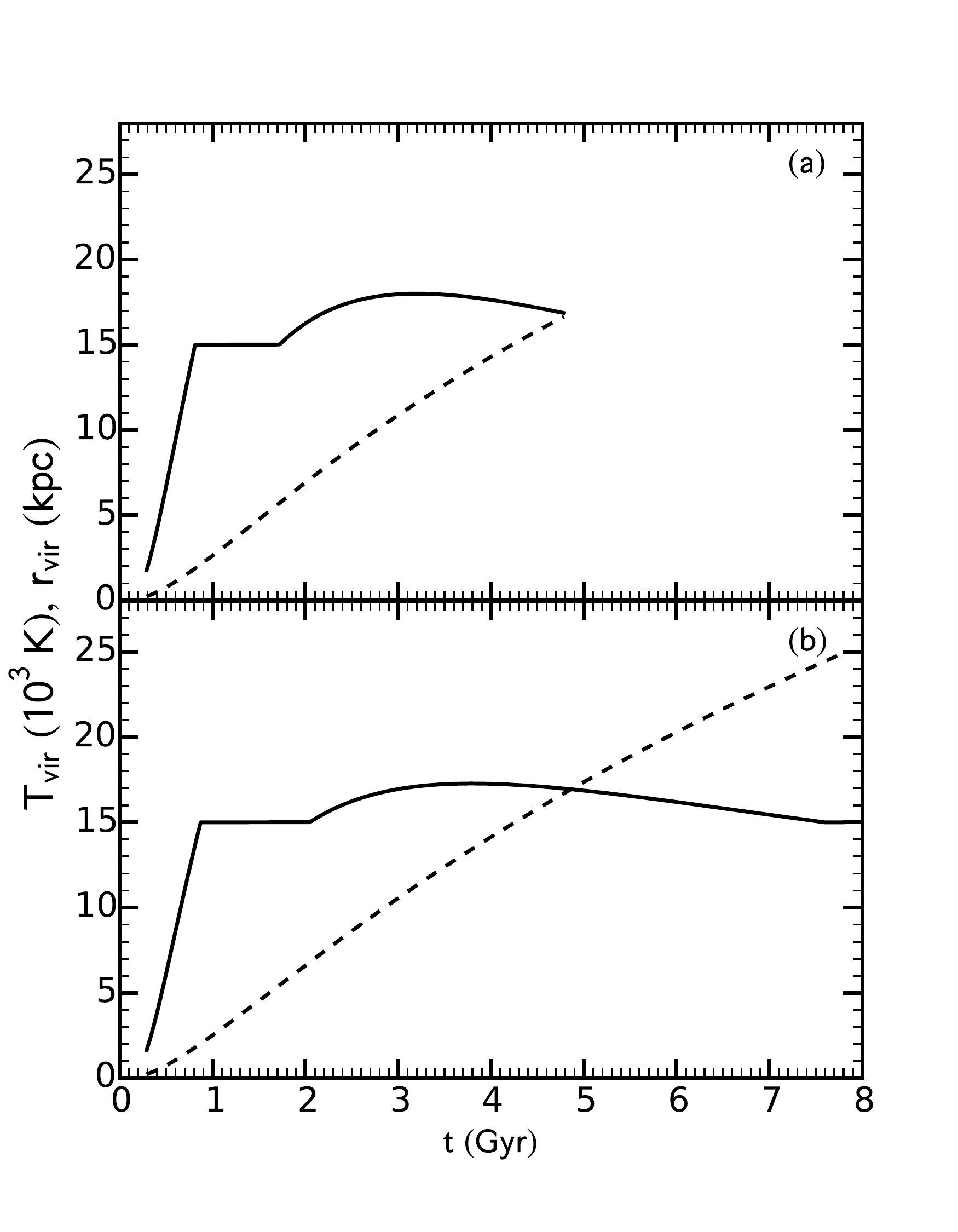}
\caption{Virial temperature $T_{\rm vir}$ (solid curve) and radius $r_{\rm vir}$ 
(dashed curve) for the Fornax halo as functions of time $t$ before it became 
an MW satellite at $t=t_{\rm sat}$. Panels (a) and (b) are for
$t_{\rm sat}\approx 4.8$ and 7.8~Gyr, respectively. The horizontal part of 
the solid curve corresponds to the transition from a neutral to a fully ionized 
primordial gas at $T_{\rm vir}=1.5\times 10^4$~K. 
See Appendix~\ref{sec-vir} for details.}
\label{fig-vir}
\end{figure}

Using the model of \cite{zhao}, we show $M_h(t)$ at $t<t_{\rm sat}$ for 
$t_{\rm sat}\approx 4.8$ and 7.8~Gyr as the solid and dashed curves,
respectively, in Fig.~\ref{fig-mh}. We now make better estimates of
$M_g(t)$ for $t<t_{\rm sat}$ by considering a growing star-forming disk.
In accordance with the model of \cite*{mo}, we assume that the effective 
disk radius grew in proportion to the virial radius $r_{\rm vir}$ of the halo 
(see Appendix~\ref{sec-vir}) until $t=t_{\rm sat}$ and then remained fixed 
during the subsequent SFH. Thus we obtain
\begin{equation}
\lambda_*(t)=\left\{\begin{array}{ll}
\lambda_*(t_{\rm sat})[r_{\rm vir}(t_{\rm sat})/r_{\rm vir}(t)]^{2(\alpha-1)},
&t< t_{\rm sat},\\
\lambda_*(t_{\rm sat}),&t\geq t_{\rm sat}.
\end{array}\right.
\label{eq-lamt}
\end{equation} 
For the baseline case of $\alpha=1.5$ and 
$\lambda_*(t_{\rm sat})=10^{-14}\,M_\odot$~yr$^{-1}$, we show
the evolution of $r_{\rm vir}$ at $t< t_{\rm sat}$ for the two estimates
of $t_{\rm sat}$ in Fig.~\ref{fig-vir}. The corresponding $M_g(t)$ 
calculated from the smooth $\psi(t)$ in Fig.~\ref{fig-sfr} is shown as the 
solid and dashed curves in Fig.~\ref{fig-mg}. 
The corresponding $\Delta F(t)$ is shown in Figs.~\ref{fig-df}b and 
\ref{fig-df}c. Note that a growing disk does not change $\Delta F(t)$ at 
$t<t_{\rm sat}$ qualitatively, and therefore, does not affect our above 
estimates of $t_{\rm sat}$.

\begin{figure}
\includegraphics[width=84mm]{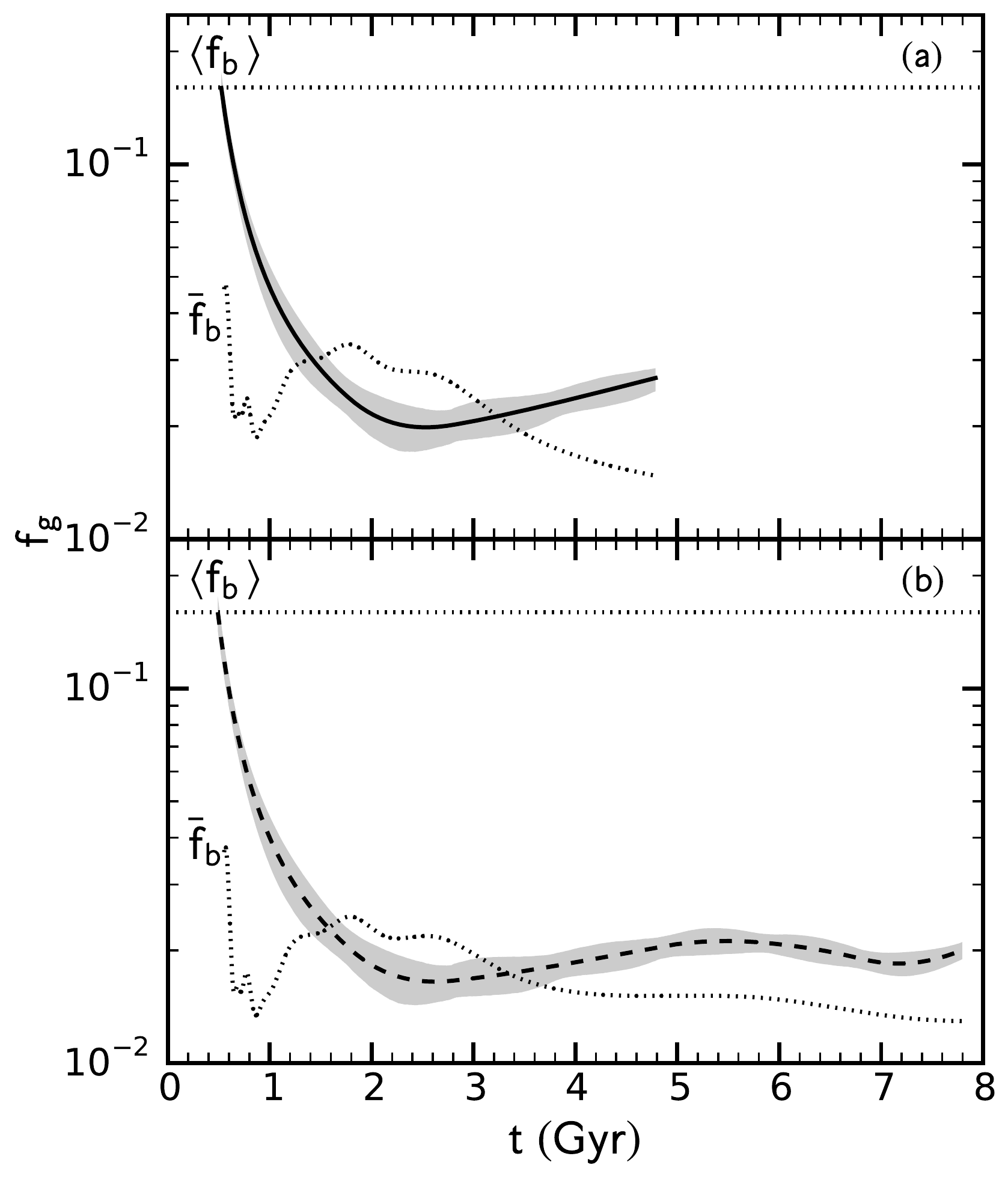}
\caption{Fornax's gas mass fraction $f_g=M_g/M_h$ as a function of time $t$ 
(solid and dashed curves) before it became an MW satellite at $t=t_{\rm sat}$ 
for the baseline case of $\alpha=1.5$ and 
$\lambda_*(t_{\rm sat})=10^{-14}\,M_\odot$~yr$^{-1}$.
Panels (a) and (b) are for $t_{\rm sat}\approx 4.8$ and 7.8~Gyr, respectively.
In each panel, the shaded region indicates the 68\% confidence interval
due to uncertainties in $M_g(t)$ only.
The dotted curve shows the mean baryon mass fraction 
$\bar f_b(t)$ calculated from Eq.~(\ref{eq-fb}) for Fornax-like halos in a 
universe reionized at $z=9$ ($t\approx 0.55$~Gyr).
The horizontal dotted line indicates the cosmic mean
mass fraction of baryons $\langle f_b\rangle$. 
We estimate $t_{\rm reion}\sim 0.52$ and 0.49~Gyr corresponding to
$z_{\rm reion}\sim 9.4$ and 9.8 for panels (a) and (b), respectively,
and set $f_g(t)=\langle f_b\rangle$ for $t< t_{\rm reion}$.
The agreement between $f_g$ and $\bar f_b$ justifies the choice of
$\lambda_*(t)$.}
\label{fig-fg}
\end{figure}

The choice of $\lambda_*(t)$ can now be justified by comparing 
our results on $M_g(t)$ and $M_h(t)$ at $t<t_{\rm sat}$
with simulations of gas accretion by halos after reionization of the 
intergalactic medium (IGM) at time $t_{\rm reion}$.
\cite*{okamoto} showed that at $t>t_{\rm reion}$, the baryonic mass fractions 
for halos of mass $M$ are distributed around the mean value
\begin{equation}
\bar f_b(M,t)=\langle f_b\rangle\{1+(2^{2/3}-1)[M_c(t)/M]^2\}^{-3/2},
\label{eq-fb}
\end{equation}
where $M_c(t)$ is a characteristic halo mass determined from their simulations
and is shown as the dotted curve in Fig.~\ref{fig-mh},
$\langle f_b\rangle\equiv\Omega_b/(\Omega_{\rm CDM}+\Omega_b)\approx 0.16$ 
is the cosmic mean mass fraction of baryons, and $\Omega_{\rm CDM}=0.243$ 
and $\Omega_b=0.047$ are the fractional contributions to the critical density of 
the present universe from CDM and baryons, respectively. The above equation 
gives $\bar f_b=\langle f_b\rangle/2$ for $M=M_c(t)$ and 
$\bar f_b=\langle f_b\rangle$ in the limit $M\gg M_c(t)$. Using $M=M_h(t)$ for
$t<t_{\rm sat}$, we calculate $\bar f_b(t)\equiv\bar f_b(M,t)$ from Eq.~(\ref{eq-fb}) 
and show the result as the dotted curve in Fig.~\ref{fig-fg}. This result can be 
compared with $f_g(t)=M_g(t)/M_h(t)$ for our baseline case with
$t_{\rm sat}\approx 4.8$ (7.8)~Gyr, which is shown 
as the solid (dashed) curve in Fig.~\ref{fig-fg}a (\ref{fig-fg}b).

In comparing the $f_g(t)$ and $\bar f_b(t)$ shown in Fig.~\ref{fig-fg}, we note
that \cite{okamoto} assumed a reionization redshift of $z=9$
corresponding to $t=0.55$~Gyr. We now estimate the time $t_{\rm reion}$
for reionization of the IGM surrounding Fornax. 
The oldest stellar age bin for the data on Fornax's SFH is 13--14~Gyr. In view of
the substantial uncertainties of $\sim 2.5$~Gyr for these ages \citep{deboer},
we take the centroid of this age bin at 13.5~Gyr as a reasonable age estimate for
Fornax's oldest stars. With an adopted age of 13.79~Gyr for the universe,
we estimate that star formation started in Fornax at 
$t_{\rm SF}\approx 0.29$~Gyr (resdhift $z_{\rm SF}\approx 14.5$), when
the halo mass reached $M_h(t_{\rm SF})\approx 1.3\times 10^6\,M_\odot$ or
$1.1\times 10^6\,M_\odot$ for $t_{\rm sat}\approx 4.8$ or 7.8~Gyr, respectively
(see Fig.~\ref{fig-mh}). This is reasonable because the corresponding virial 
temperature $T_{\rm vir}(t_{\rm SF})\approx 1.8\times 10^3$ or
$1.6\times 10^3$~K (see Appendix~\ref{sec-vir} and Fig.~\ref{fig-vir}) could 
readily enable cooling by H$_2$ molecules as required for formation of the 
first stars (see e.g., \citealt{bromm} for a review). 
With $t_{\rm SF}\approx 0.29$~Gyr, star formation most likely started in Fornax 
at $t< t_{\rm reion}$. For $t_{\rm SF}< t<t_{\rm reion}$, accretion of gas by 
the Fornax halo should be proportional to that of CDM and $f_g(t)=M_g(t)/M_h(t)$ 
should be close to the cosmic mean mass fraction of baryons 
$\langle f_b\rangle\approx 0.16$. The $f_g(t)$ shown in Fig.~\ref{fig-fg}a 
(\ref{fig-fg}b) reaches $\langle f_b\rangle$ at $t\approx 0.52$ (0.49)~Gyr and 
exceeds this value at earlier times (not shown). Our assumed SFL most likely
does not apply to the early epoch near the onset of star formation. So we
could take $t_{\rm reion}\sim 0.52$ (0.49)~Gyr corresponding to 
$z_{\rm reion}\sim 9.4$ (9.8) and make a revised estimate of
$ M_g(t)\approx\langle f_b\rangle M_h(t)$ for $t_{\rm SF}< t<t_{\rm reion}$.
The above estimates of $z_{\rm reion}$ are in good agreement with 
$z=8^{+3}_{-2}$ found for reionization of the IGM surrounding MW dwarf 
galaxies in simulations by \cite{busha}.

With $z_{\rm reion}\sim 9.4$ (9.8) being close to the value assumed by
\cite{okamoto}, our estimated $f_g$ can be directly compared with
$\bar f_b$ and the agreement is rather good as shown in Fig.~\ref{fig-fg}.
We note that this agreement is not affected by adding the contribution of 
stars to $f_g$ as the mass of stars is always significantly smaller 
than that of gas for $t<t_{\rm sat}$.  
Therefore, we consider that our choice of $\lambda_*(t)$ with
$\lambda_*(t_{\rm sat})=10^{-14}\,M_\odot$~yr$^{-1}$ for the baseline 
case is reasonable.

\subsection{Other cases}
\label{sec-a12}
\begin{figure}
\includegraphics[width=84mm]{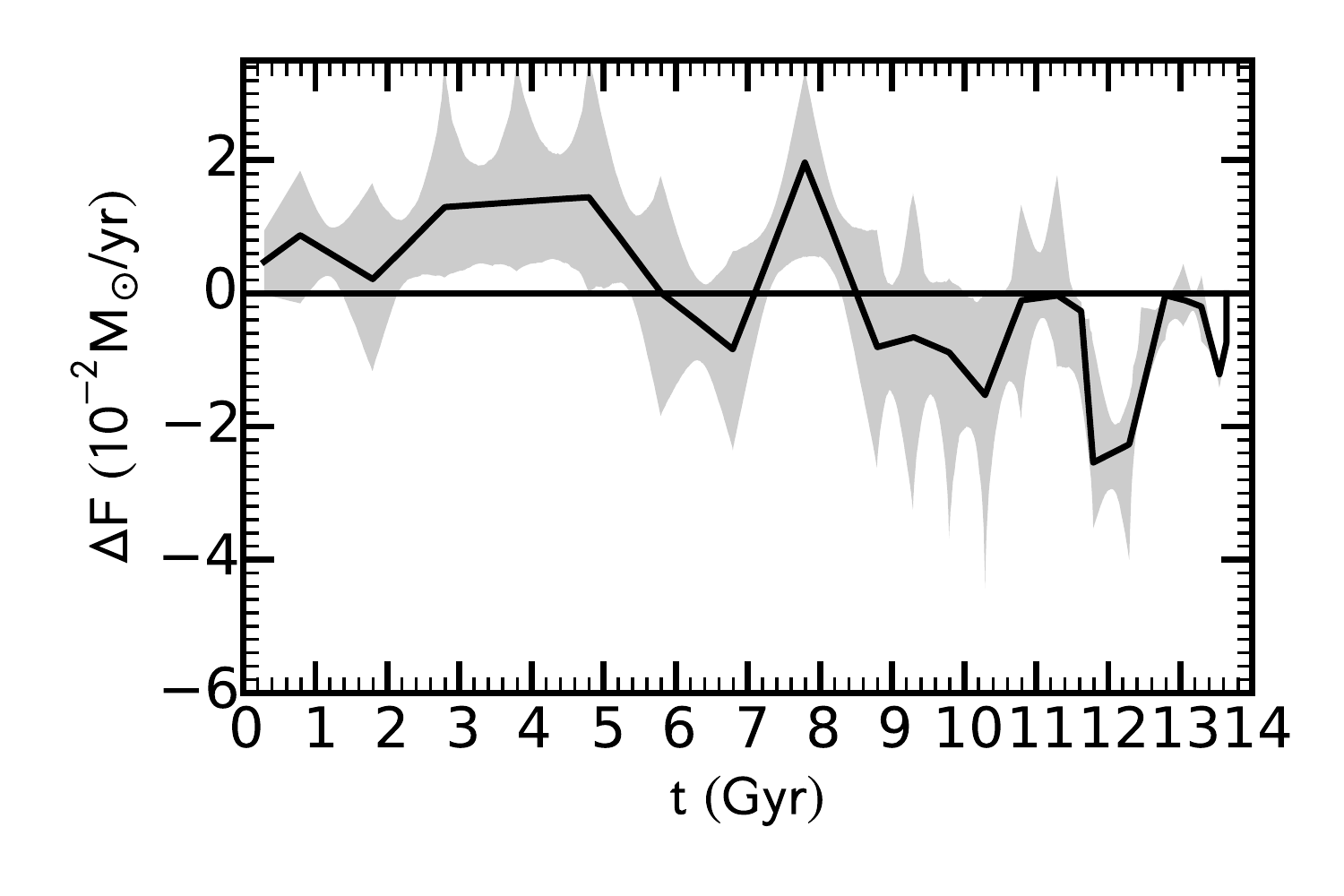}
\caption{Same as Fig.~\ref{fig-df}a but for the case of $\alpha=1$ 
and $\lambda_*(t_{\rm sat})=7\times 10^{-11}\,M_\odot$~yr$^{-1}$.
Note that $\lambda_*(t)=\lambda_*(t_{\rm sat})$ is independent of 
time for this case.}
\label{fig-df1}
\end{figure}

\begin{figure}
\includegraphics[width=84mm]{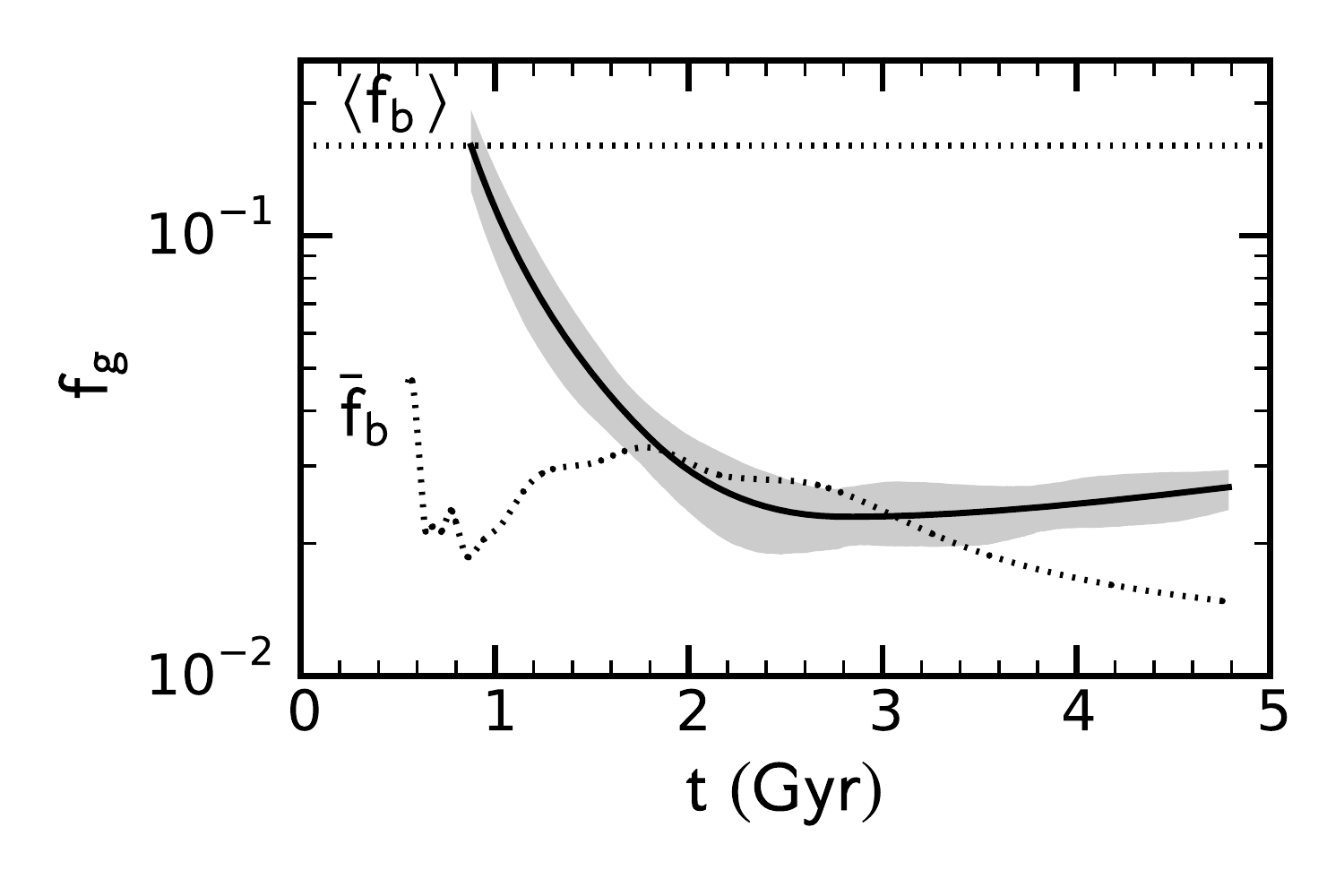}
\caption{Same as Fig.~\ref{fig-fg}a but for the case of $\alpha=1$ and 
$\lambda_*(t_{\rm sat})=7\times 10^{-11}\,M_\odot$~yr$^{-1}$. Note
that $\lambda_*(t)=\lambda_*(t_{\rm sat})$ is independent of time
for this case.}
\label{fig-fg1}
\end{figure}

The case of $\alpha=1$ is special because 
$\lambda_*(t)=\lambda_*(t_{\rm sat})$ is independent of time
(see \S\ref{sec-sflfor}). We choose $\lambda_*(t_{\rm sat})$ 
for this case to obtain the same $M_g$ at $t\approx 4.8$~Gyr 
as the top curve in Fig.~\ref{fig-mg} for the baseline case. This gives
$\lambda_*(t_{\rm sat})=7\times 10^{-11}\,M_\odot$~yr$^{-1}$,
which is consistent with the estimate based on empirical SFLs 
as discussed in \S\ref{sec-sflfor}. The corresponding
$\Delta F(t)$ is shown in Fig.~\ref{fig-df1}, where the same
onset of suppression of the net gas inflow at 
$t_{\rm sat}\approx 4.8$ or 7.8~Gyr as for the baseline case
can be seen. In Fig.~\ref{fig-fg1} we take $t_{\rm sat}\approx 4.8$~Gyr
and show the corresponding $f_g(t)$, which gives
$t_{\rm reion}\sim 0.88$~Gyr ($z_{\rm reion}\sim 6.3$).
As $\bar f_b$ assumes $z=9$ for reionization, straightforward
comparison between $f_g$ and $\bar f_b$ can be made only 
for $t>0.88$~Gyr. The agreement is rather good for $t\ga 1.2$~Gyr.

\begin{figure}
\includegraphics[width=84mm]{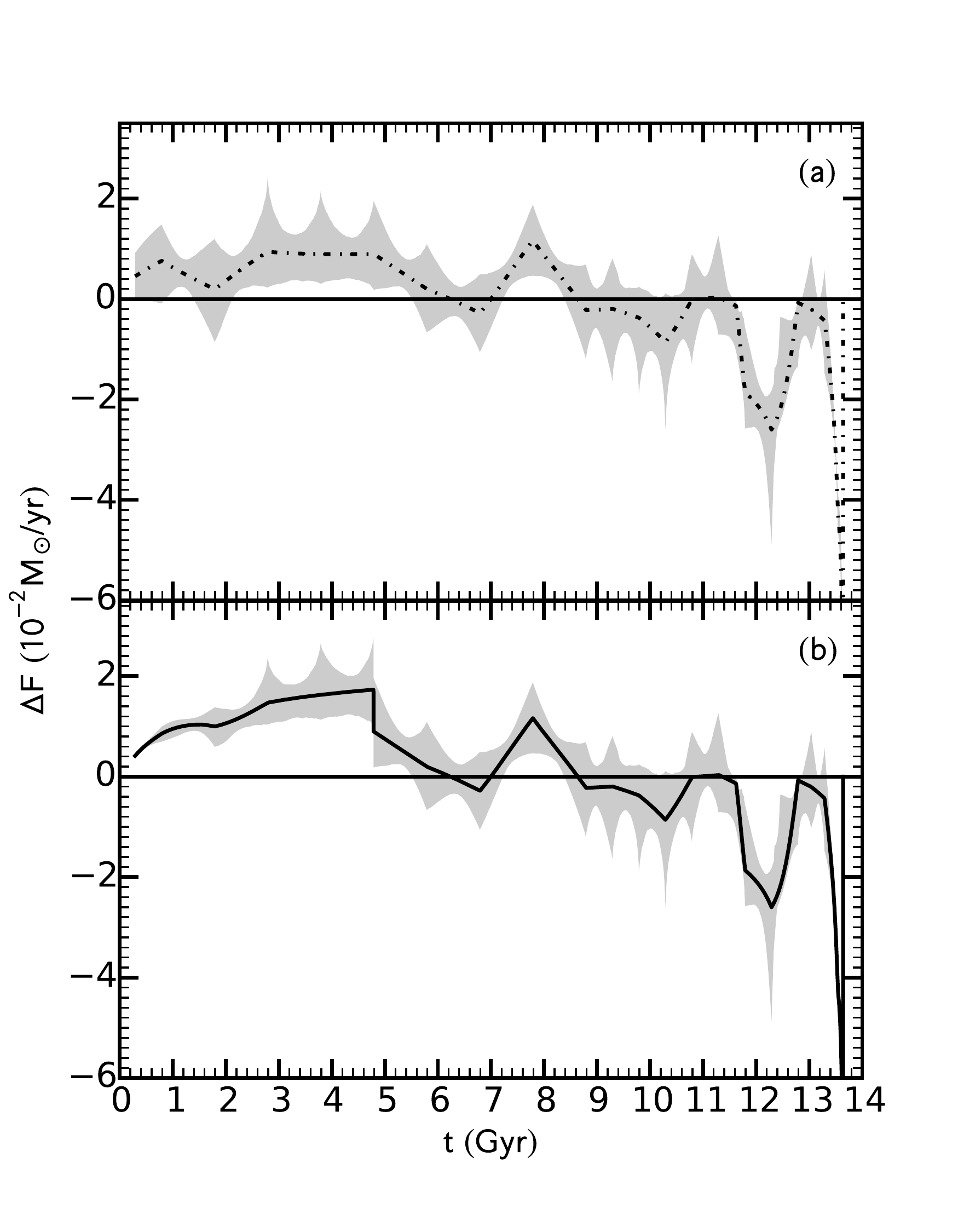}
\caption{Same as Figs.~\ref{fig-df}a and \ref{fig-df}b 
but for the case of $\alpha=2$ and 
$\lambda_*(t_{\rm sat})=1.4\times 10^{-18}\,M_\odot$~yr$^{-1}$.}
\label{fig-df2}
\end{figure}

\begin{figure}
\includegraphics[width=84mm]{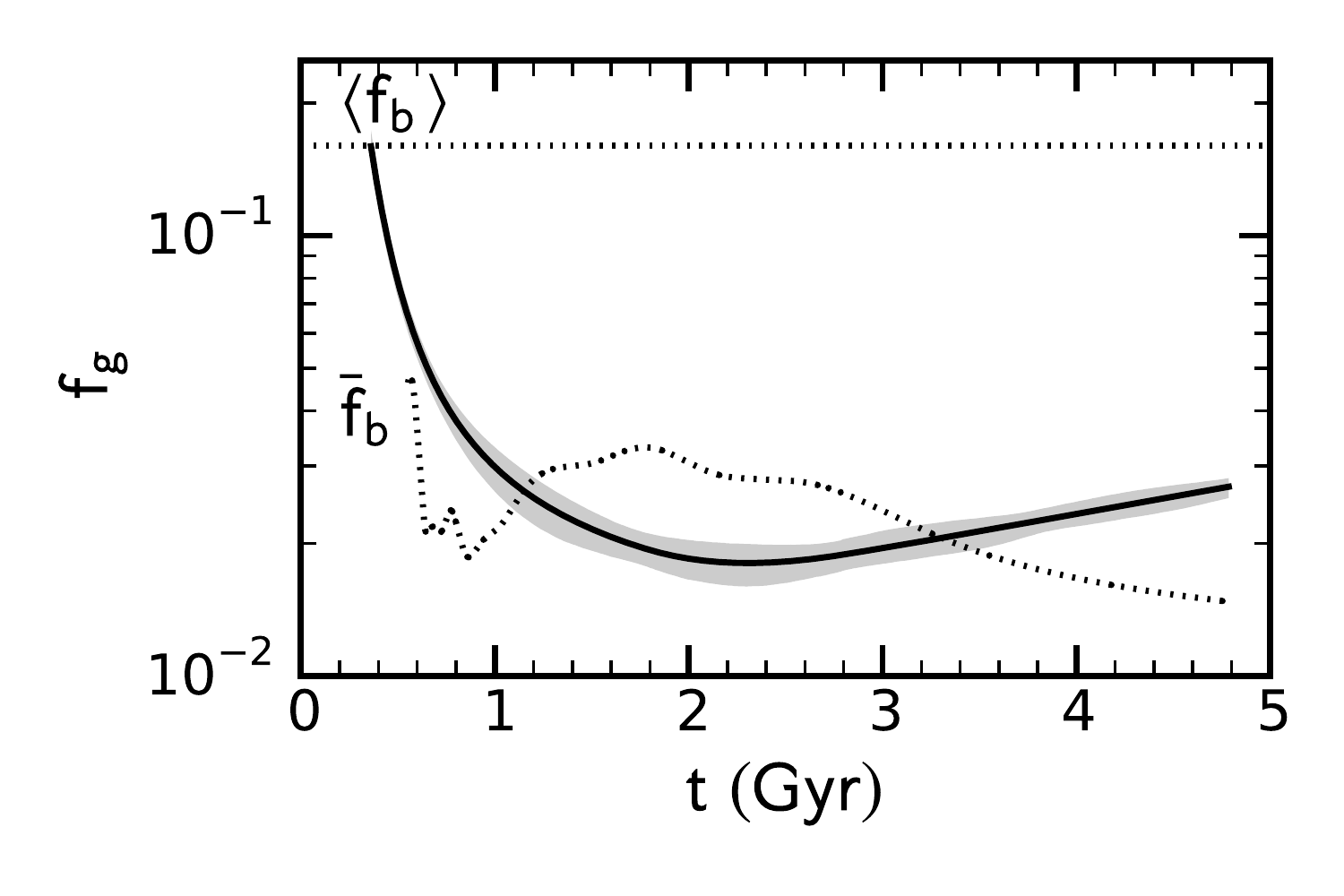}
\caption{Same as Fig.~\ref{fig-fg}a but for the case of $\alpha=2$ and 
$\lambda_*(t_{\rm sat})=1.4\times 10^{-18}\,M_\odot$~yr$^{-1}$.}
\label{fig-fg2}
\end{figure}

For the case of $\alpha=2$, we first consider
$\lambda_*(t)=\lambda_*(t_{\rm sat})$ and choose
$\lambda_*(t_{\rm sat})$ to obtain the same $M_g$ at 
$t\approx 4.8$~Gyr as the top curve in Fig.~\ref{fig-mg} 
for the baseline case. This gives
$\lambda_*(t_{\rm sat})=1.4\times 10^{-18}\,M_\odot$~yr$^{-1}$ 
in accord with the estimate based on empirical SFLs 
(see \S\ref{sec-sflfor}). The corresponding $\Delta F(t)$ shown
in Fig.~\ref{fig-df2}a again suggests $t_{\rm sat}\approx 4.8$ or 
7.8~Gyr. The $\Delta F(t)$ for a disk growing until $t\approx 4.8$~Gyr
is shown in Fig.~\ref{fig-df2}b and the corresponding $f_g(t)$ in 
Fig.~\ref{fig-fg2}. An estimate of $t_{\rm reion}\sim 0.36$~Gyr 
($z_{\rm reion}\sim 12.3$) can be made and there is good
agreement between $f_g$ and $\bar f_b$ at $t>0.55$~Gyr, i.e.,
after the time of reionization assumed for the latter.

Based on the above results, we consider that our results on
$t_{\rm sat}$, and hence $M_h(t_{\rm sat})$, are insensitive to 
the exact form of our assumed SFL so long as it is consistent 
with the empirical SFLs discussed in \S\ref{sec-sflfor}.

\section{Global Gas Dynamics and Evolution of Fornax as an MW Satellite}
\label{sec-sat}
\begin{figure}
\includegraphics[width=84mm]{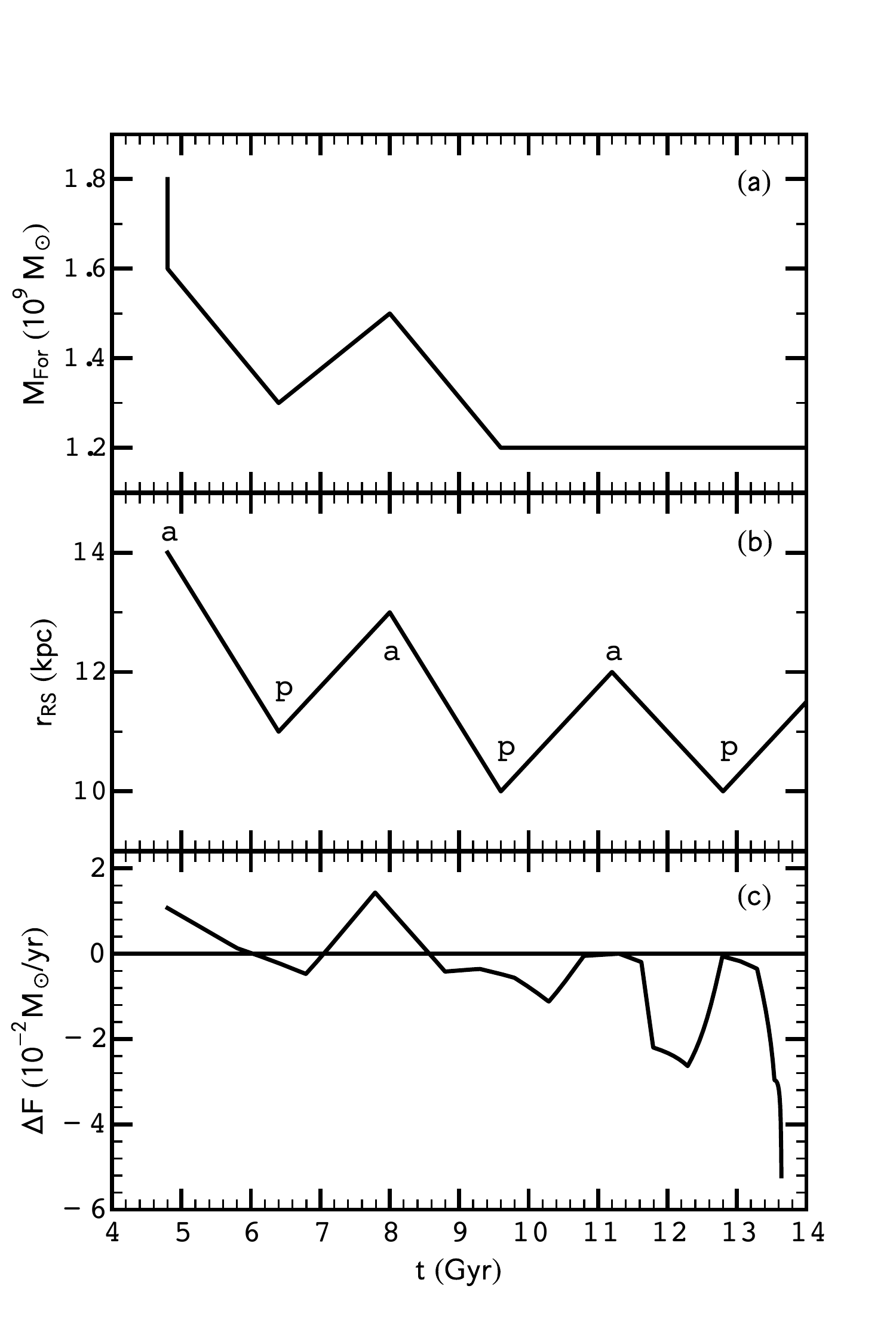}
\caption{Evolution of the Fornax halo as an MW satellite at $t> 4.8$~Gyr. 
Its total mass $M_{\rm For}$ (a) and Roche-sphere radius $r_{\rm RS}$ 
(b) are estimated at the apogalacticon (indicated by ``a'') and 
perigalacticon (indicated by ``p''), and connected by line segments to 
guide the eye. For convenience of discussing the effects of tidal interaction, 
the relevant $\Delta F(t)$ for the baseline case (c) is also shown.
See text for details.}
\label{fig-tide}
\end{figure}

Based on the discussions in \S\ref{sec-gas}, the baseline case appears
to give representative results on the general evolution of Fornax. 
Below we focus on this case with $t_{\rm sat}\approx 4.8$~Gyr to 
discuss the global gas dynamics and evolution of Fornax as an MW Satellite. 
Considering that our discussion is mostly qualitative, we use the solid curve 
for $\Delta F(t)$ at $t>4.8$~Gyr in Fig.~\ref{fig-df}b as a guide. This is
reproduced in Figure~\ref{fig-tide}c, which shows that $\Delta F$ dropped 
to zero at $t\approx 6$~Gyr and the first net outflow ($\Delta F<0$) occurred 
for $t\approx 6$--7~Gyr. This was followed by the final round of net inflow 
($\Delta F>0$) for $t\approx 7$--8.6~Gyr, and then by several episodes of net 
outflow until star formation ended in Fornax. As we discuss below, the episodic 
gas flows at $t>4.8$~Gyr shown in Fig.~\ref{fig-tide}c were driven by Fornax's
orbital motion and its tidal interaction with the MW.

\subsection{Orbital motion and tidal interaction}
\label{sec-orb}
Observations show that Fornax currently moves in an elliptical orbit with a period of 
$\sim 3.2$~Gyr. Its distance from the MW center is $R_p\sim 118$~kpc at the 
perigalacticon and $R_a\sim 152$~kpc at the apogalacticon \citep{piatek}. Using the 
median halo growth history and the corresponding halo structure given by the model 
of \cite{zhao}, we find that for a present-day MW 
halo mass in the range of (1--$2)\times 10^{12}\,M_\odot$, the MW mass 
enclosed by Fornax's present-day orbit is $\sim 1.5$ times the MW halo mass
at $t_{\rm sat}\approx 4.8$~Gyr when Fornax became a satellite. In view
of this moderate change, we assume that Fornax's orbit for $t\ga 4.8$~Gyr 
can be approximately described by the present-day parameters.
For estimates, we take the MW mass enclosed within Fornax's orbit to be
$M_{\rm MW}^{\rm encl}\sim 7.5\times 10^{11}\,M_\odot$ on average.
This gives an average orbital velocity of
$v_{\rm For}\sim (GM_{\rm MW}^{\rm encl}/a)^{1/2}\sim150$~km/s for Fornax, where 
$a=(R_p+R_a)/2\sim 135$~kpc. For comparison, the circular velocity characterizing 
matter motion inside the Fornax halo at $t_{\rm sat}\approx 4.8$~Gyr is
$v_{\rm circ}=(GM_h/r_{\rm vir})^{1/2}\approx 21$~km/s with
$M_h\approx 1.8\times 10^9\,M_\odot$ and $r_{\rm vir}\approx 17$~kpc
(see Fig.~\ref{fig-vir}). 

The effects of tidal interaction between the MW and Fornax can be estimated
from the radii $r_{{\rm RS},p}$ and $r_{{\rm RS},a}$ of the instantaneous 
Roche spheres \citep{king} at the perigalacticon and apogalacticon, respectively. 
Ignoring the small eccentricity of $\sim 0.13$ for Fornax's orbit \citep{piatek}, 
we have
\begin{equation}
\frac{r_{{\rm RS},p}}{R_p}\approx\frac{r_{{\rm RS},a}}{R_a}
\approx\left(\frac{M_{\rm For}}{3M_{\rm MW}^{\rm encl}}\right)^{1/3},
\label{eq-roche}
\end{equation}
where $M_{\rm For}$ is the total mass of Fornax and generally differs from
$M_h$ at $t_{\rm sat}\approx 4.8$~Gyr due to tidal interaction.
The CDM and gas outside the Roche sphere would be lost from Fornax. 
However, because the Roche sphere would 
expand as Fornax moved from the perigalacticon to the apogalacticon, some of
the CDM and gas lost earlier could be reaccreted. Guided by the simulations of
\cite{han} and our derived $\Delta F$, we assume that reaccretion occurred 
only once. Subsequently, $M_{\rm For}$ settled down to a value set by the mass 
enclosed within $r_{{\rm RS},p}$. For estimates, we use the density profile given
by the model of \cite{zhao} for a halo with $M_h\approx 1.8\times 10^9\,M_\odot$ 
at $t_{\rm sat}\approx 4.8$~Gyr. Assuming that Fornax was at the apogalacticon 
at $t_{\rm sat}\approx 4.8$~Gyr with 
$M_{\rm For}\approx 1.8\times 10^9\,M_\odot$, we estimate that its total mass was
quickly reduced to $\sim 1.6\times 10^9\,M_\odot$, which corresponds to the mass 
enclosed within the Roche sphere at this time. The mass of the Fornax halo was
further reduced to $\sim 1.3\times 10^9\,M_\odot$ at $t\sim 6.4$~Gyr when it was 
at the perigalacticon with a smaller Roche sphere. It then reaccreted some of the 
CDM (and gas) lost earlier to obtain $M_{\rm For}\sim 1.5\times 10^9\,M_\odot$ 
at $t\sim 8$~Gyr when it was at the apogalacticon again. After that $M_{\rm For}$ 
settled down to $\sim 1.2\times 10^9\,M_\odot$, which is enclosed within 
$r_{{\rm RS},p}$ estimated from Eq.~(\ref{eq-roche}) for this value of $M_{\rm For}$.
During the above evolution of the Fornax halo, $r_{{\rm RS},a}$ started at 
$\sim 14$~kpc and settled down to $\sim 12$~kpc while $r_{{\rm RS},p}$ 
decreased from $\sim 11$~kpc to $\sim 10$~kpc. The evolution of $M_{\rm For}$, 
$r_{{\rm RS},a}$, and $r_{{\rm RS},p}$ is sketched in Fig.~\ref{fig-tide}.

Based on the above discussion, we estimate that Fornax lost $\sim 1/3$ of its 
CDM through tidal interaction. However, this loss occurred in the outermost
region of the halo and had little effect on the structure of its interior. For example,
because $r_{1/2}\approx 944$~pc was well within the Roche sphere at all times,
the mass $M(<r_{1/2})$ enclosed within $r_{1/2}$, which is used to 
determine $M_h$ at $t_{\rm sat}\approx 4.8$~Gyr (see Appendix~\ref{sec-mh}),
would not have been affected by tidal interaction. Similarly, with a radius of 
$\sim 2$~kpc, the star-forming disk would have remained largely intact as Fornax
orbited the MW so long as there was a sufficient supply of cold gas.

\subsection{Gas inflow and outflow}
\label{sec-flow}
Over Fornax's SFH, the star-forming disk was surrounded by hot gas either accreted
from the IGM or expelled from the disk due to feedback from star formation, which
includes SN explosions and heating by stellar radiation. Cooling of this hot gas gave 
rise to an inflow onto the disk while stellar feedback produced an outflow from it. The 
net gas flow rate $\Delta F$ represents the difference between the two.
At $t<4.8$~Gyr, accretion of the IGM by the growing Fornax halo dominated its global 
gas dynamics and supplied hot gas that cooled to provide a net inflow. 
In general, reducing 
the amount of hot gas would suppress the inflow, thereby causing $\Delta F$ to 
decrease. This accounts for the decline of $\Delta F$ for $t\approx 4.8$--6~Gyr 
(see Fig.~\ref{fig-tide}) as accretion of the IGM was disrupted and the initial tidal 
interaction between Fornax and the MW resulted in loss of hot gas along with CDM.
This gas loss was also aided by ram-pressure stripping as discussed in \S\ref{sec-ram}.
The corresponding suppression of the inflow allowed the brief first occurrence 
of a net outflow for $t\approx 6$--7~Gyr as hot gas escaped from the star-forming 
disk. This outflow carried only $\sim 2\times 10^6\,M_\odot$ and was soon quenched 
when $\sim 10^7\,M_\odot$ of the hot gas lost earlier was reaccreted and cooled to
produce an inflow for $t\approx 7$--8.6~Gyr (see Fig.~\ref{fig-tide}). 

As mentioned in \S\ref{sec-orb}, some of the gas and CDM lost earlier was reaccreted
because the Roche sphere expanded at the apogalacticon (see Fig.~\ref{fig-tide}).
This reaccretion was discussed for gas by \cite*{nichols} and seen for CDM in simulations 
by \cite{han}. However, our derived $\Delta F$ indicates that there was only one episode
of significant reaccretion for Fornax, which might represent a variation from 
the scenario of episodic starbursts with repeated gas expulsion and reaccretion 
as discussed by \cite{nichols}. In the absence of further reaccretion, several episodes
of net outflow occurred in Fornax for $t\approx 8.6$--13.6~Gyr (see Fig.~\ref{fig-tide}).
We consider that the episodic nature of these outflows is mainly due to Fornax's orbital 
motion and is very different from the episodic behavior purely driven by stellar feedback
as discussed for example, by \cite{stinson}. Specifically, we argue that although stellar
feedback initiated the escape of hot gas from the star-forming disk, the loss of such gas
from the Fornax halo at $t\approx 8.6$--13.6~Gyr was driven by tidal interaction and 
ram-pressure stripping. 

\subsection{Ram-pressure stripping}
\label{sec-ram}
We consider that the gas subject to ram-pressure stripping was dispersed 
outside Fornax's star-forming disk but within its Roche sphere of $\sim 10$~kpc 
in radius. From the $\Delta F$ shown in Fig~\ref{fig-tide}c, we estimate that 
$\sim 10^7\,M_\odot$ of gas were stripped for each episode. So this gas
had a density of $n\sim 10^{-4}$~cm$^{-3}$ while inside Fornax. Its
internal pressure was $P\sim nkT\sim m_pnv_{\rm th}^2$, with
$T$ being its temperature, 
$v_{\rm th}\sim(kT/m_p)^{1/2}\sim 29(T/10^5\ {\rm K})^{1/2}$~km/s
its thermal velocity, and $k$ the Boltzmann constant. The ram pressure can be
estimated as
\begin{equation}
P_{\rm ram}\sim m_pn_{\rm med}v_{\rm For}^2,
\end{equation} 
where $n_{\rm med}$ is the density of the gaseous medium through which
Fornax moved. Ram-pressure stripping occurred when $P_{\rm ram}$ exceeded 
$P$, i.e.,
\begin{equation}
n_{\rm med}v_{\rm For}^2>nv_{\rm th}^2.
\end{equation}
For the hot gas accreted from the reionized IGM right before Fornax became 
an MW satellite, we estimate $v_{\rm th}\sim v_{\rm circ}\sim 21$~km/s.
This gas would be stripped for $n_{\rm med}\ga 2\times 10^{-6}$~cm$^{-3}$.

The majority of the gas stripped from Fornax first escaped from 
its star-forming disk due to stellar feedback. For a crude estimate, we assume
that the gas in this disk consisted of three components \citep{mckee}:
cold gas with a density of $n_{\rm cold}\sim 40$~cm$^{-3}$ and a temperature of 
$T_{\rm cold}\sim 80$~K, warm partially-ionized gas with 
$n_{\rm warm}\sim 0.3$~cm$^{-3}$ and $T_{\rm warm}\sim 8000$~K, and 
hot ionized gas with $n_{\rm hot}\sim 4\times 10^{-3}$~cm$^{-3}$ and 
$T_{\rm hot}\sim 5\times 10^5$~K. The total gas mass was dominated by the 
cold component that fed star formation. Cold gas was also converted by stellar 
feedback to maintain the hot component as hot gas escaped from the disk in an 
outflow. Assuming adiabatic cooling during its escape, we estimate an upper limit 
of $T\sim T_{\rm hot}(n/n_{\rm hot})^{2/3}\sim4\times 10^4$~K corresponding to
$v_{\rm th}\sim 18$~km/s for the hot gas when it was dispersed inside the Roche 
sphere. Its stripping requires $n_{\rm med}\ga 10^{-6}$~cm$^{-3}$.

Based on the above discussion, with $n_{\rm med}\ga 2\times 10^{-6}$~cm$^{-3}$
for the gaseous medium around Fornax's orbit at $\sim 118$--152~kpc from the 
MW center, ram-pressure stripping would have occurred to both accreted hot gas 
for $t\approx 4.8$--6~Gyr and gas heated by stellar feedback for $t\approx 8.6$--13.6~Gyr
in Fornax. The gas density in the present MW halo can only be 
estimated by indirect means and its time evolution over the MW's history can only be 
estimated by simulations of galaxy formation. Simulations of \cite{kauf} give
$n_{\rm med}\sim 10^{-5}$--$10^{-4}$~cm$^{-3}$ at $\sim 100$--200~kpc from the 
galactic center with no significant changes over a period of 10~Gyr. These results are
consistent with estimates based on observations of MW halo clouds (e.g., \citealt{hsu}).
However, \cite{murali} derived a present limit of 
$n_{\rm med}< 10^{-5}$~cm$^{-3}$ at 50~kpc from the MW center by considering 
effects on the Magellanic Stream. For comparison, the cosmic mean baryon density 
at $t_{\rm sat}\approx 4.8$~Gyr ($z_{\rm sat}\approx 1.33$) was 
$\langle n_b\rangle\approx 3\times 10^{-6}$~cm$^{-3}$.
We consider $n_{\rm med}\sim 3\times 10^{-6}$--$10^{-5}$~cm$^{-3}$ as reasonable
around Fornax's orbit for $t\ga 4.8$~Gyr. Therefore, ram-pressure stripping would
have been effective in removing gas from Fornax. 

For the three gas components in Fornax's star-forming disk with densities and 
temperatures assumed above, they were in pressure equilibrium and their internal 
pressure was $\ga 10^2$ times the ram pressure. Consequently, for those conditions
the ram pressure would have had little impact on the disk. However, when a sufficient
amount of gas had been removed, the three-component structure would be disrupted
and gas density in the disk would drop, eventually allowing gas to be ram-pressure 
stripped directly from the disk. This might account for the sharp increase in the
net outflow rate at the end ($t\sim 12.8$--13.6~Gyr, see Fig~\ref{fig-tide}c).

\subsection{Gas loss from Fornax}
In summary, we propose the following outline for gas dynamics in Fornax
as it orbited the MW. Tidal interaction was responsible for removing $\sim 1/3$ of
its original CDM. Because collisionless CDM does not suffer ram pressure,
the CDM lost during Fornax's first orbital period appeared to have stayed close
to the orbit and kept the simultaneously-lost gas from dispersing.
This was crucial to the reaccretion of some of the lost CDM and gas as the Roche 
sphere expanded near the end of this period. Subsequently, there was no significant
clustering of CDM near the orbit and once ram-pressure-stripped gas moved 
outside the largest Roche sphere of $\sim 12$~kpc in radius, it was lost from 
Fornax. We note that although the change in the Roche-sphere radius from
$\sim 10$~kpc at the perigalacticon to $\sim 12$~kpc at the apogalacticon
appeared small, the corresponding change in the Roche-sphere volume was by 
a significant factor of $\sim 1.7$. The non-monotonic evolution of net gas outflow for 
$t\approx 8.6$--13.6~Gyr might have resulted because an expanding 
Roche sphere would oppose while a shrinking one would enhance
the effect of ram pressure. This would explain the coincidence of 
$\Delta F\sim 0$ at $t\sim 11.2$~Gyr with the transition from an expanding 
Roche sphere to a shrinking one (Fig.~\ref{fig-tide}). We also note that subsequently
the net outflow rate $|\Delta F|$ increased as Fornax approached its perigalacticon. 
However, the net outflow rate started to decrease at $t\sim 12.3$~Gyr and dropped 
to $\sim 0$ when Fornax reached its perigalacticon at $t\sim 12.8$~Gyr. This is
in contradiction to the expected effect of the smallest Roche sphere at the
perigalacticon. Interestingly, several observations 
(e.g., \citealt{col2004,col2005,deboer4})
found evidence that Fornax might have accreted a smaller satellite or
reaccreted some of the lost gas $\sim 1.5$--2~Gyr ago
($t\sim 11.8$--12.3~Gyr). This would have explained the puzzling drop of
the net outflow rate described above. In any case, it appears that Fornax 
lost all of its gas at $t\approx 13.6$~Gyr before returning to its apogalacticon
(estimated to occur at $t\sim 14.4$~Gyr or $\sim 0.6$~Gyr from now).

Gas loss from satellites through ram-pressure stripping and tidal interaction
was studied in detail by simulations of \cite{mayer}, who emphasized the
importance of the UV background in keeping the gas widely distributed inside
the satellites. Gas depletion in Local Group dwarfs was studied analytically
by \cite{nichols11}, who emphasized heating of gas by stellar feedback.
In general accord with these detailed studies, we propose that gas in the 
star-forming disk was first heated by SNe to $T_{\rm hot}\sim 5\times 10^5$~K,
and then escaped to be dispersed outside the disk before getting
removed from Fornax. It is also possible that heating by both SNe and the UV 
background was needed, but we focus on SN heating below.

Using the same initial mass function as adopted by 
\cite{deboer} in deriving the SFR and assuming that stars of 8--$120\,M_\odot$ 
result in SNe, we obtain an SN rate of $R_{\rm SN}(t)\approx10^{-2}\psi(t)/M_\odot$.
For SNe to heat a total mass $\Delta M_{\rm hot}$ of gas that escaped during
an episode of outflow, the required efficiency per SN is
\begin{eqnarray}
\epsilon_{\rm SN}&\sim&\frac{(\Delta M_{\rm hot}/m_p)kT_{\rm hot}}
{\Delta N_{\rm SN}E_{\rm expl}}\nonumber\\&\sim&0.083
\left(\frac{\Delta M_{\rm hot}}{10^7\,M_\odot}\right)
\left(\frac{T_{\rm hot}}{5\times 10^5\ {\rm K}}\right)
\left(\frac{10^4}{\Delta N_{\rm SN}}\right)\nonumber\\
&&\times\left(\frac{10^{51}\ {\rm ergs}}{E_{\rm expl}}\right),
\end{eqnarray}
where $\Delta N_{\rm SN}$ is the total number of SNe occurring in this episode 
and $E_{\rm expl}$ is the average explosion energy per SN.
From the $\psi$ and $\Delta F$ shown in Figs.~\ref{fig-sfr} and \ref{fig-tide}, 
respectively, we estimate that $\Delta M_{\rm hot}\sim 1.2\times 10^7\,M_\odot$
($2.2\times 10^7\,M_\odot$) and $\Delta N_{\rm SN}\approx 6\times 10^4$
($1.4\times 10^4$) for the episode of $t\approx 8.6$--11.2~Gyr (11.2--12.8~Gyr),
which correspond to $\epsilon_{\rm SN}\sim 1.7$\% (13\%) for 
$T_{\rm hot}\sim 5\times 10^5$~K and $E_{\rm expl}\sim 10^{51}$~ergs. 
For the last episode of $t\approx 12.8$--13.6~Gyr, there was a sharp increase in 
the net outflow rate and $\Delta M_{\rm hot}\sim 8\times 10^6\,M_\odot$,
which would have required $\epsilon_{\rm SN}\sim 51$\% with
$\Delta N_{\rm SN}\approx 1.3\times 10^3$.
As noted in \S\ref{sec-ram}, for this terminal episode, gas could have
been ram-pressure stripped directly from the star-forming disk due to the decrease
in density. So a lower SN-heating efficiency could have been sufficient. 
Heating by Type Ia SNe, which is ignored in the above discussion,
could also reduce the required $\epsilon_{\rm SN}$.
Therefore, it appears that SN-heating coupled with ram-pressure stripping and 
tidal interaction for a few orbital periods was responsible for removing gas 
and terminating star formation in Fornax.

\section{Discussion and Conclusions}
\label{sec-dc}
We have presented an empirical model for the halo evolution and global gas dynamics
of Fornax. Guided by data on star formation, especially those reported by \cite{bigiel} 
and \cite{roy} for nearby galaxies, we have assumed a global SFR 
$\psi(t)=\lambda_*(t)[M_g(t)/M_\odot]^\alpha$ that is dependent on the total mass 
$M_g(t)$ for cold gas in the star-forming disk. We have examined three different cases 
of $\alpha=1$, 1.5, and 2, and chosen the corresponding $\lambda_*(t)$ in agreement with 
the empirical SFLs. For each case, we have used the data on Fornax's $\psi(t)$ 
provided by \cite{deboer} to derive $M_g(t)$ and estimate the rate of net gas flow 
$\Delta F(t)$ to or from Fornax's star-forming disk. We have identified the onset of the 
transition in $\Delta F(t)$ from a net inflow to a net outflow as the time $t_{\rm sat}$ 
at which Fornax ceased evolving independently and became an MW satellite. 
We have determined the mass $M_h(t_{\rm sat})$ of the Fornax halo at this time by 
using the median halo growth history in the model of \cite{zhao} and requiring that 
the corresponding density profile give the mass enclosed within the half-light radius 
as derived from observations. We have further justified our chosen $\lambda_*(t)$ by
comparing the gas mass fraction $f_g(t)=M_g(t)/M_h(t)$ at $t<t_{\rm sat}$ with the 
results from cosmological simulations of \cite{okamoto} on gas accretion by halos 
in a reionized universe. 

Our main results for Fornax are the evolution of its total gas mass $M_g(t)$ and 
net gas flow rate $\Delta F(t)$, the time $t_{\rm sat}$ when it became an MW satellite, 
and its halo mass $M_h(t_{\rm sat})$ at this time. For our baseline case of 
$\alpha=1.5$ and $\lambda_*(t_{\rm sat})=10^{-14}\,M_\odot$~yr$^{-1}$, we have 
obtained $t_{\rm sat}\approx 4.8$ (7.8)~Gyr and 
$M_h(t_{\rm sat})\approx 1.8\times 10^9\,M_\odot$ ($2.4\times 10^9\,M_\odot$).
These results are in broad agreement with previous studies based on Fornax's
orbital motion \citep{rocha} and stellar kinematics \citep{strig}.
We also found that the evolution of the gas mass fraction $f_g(t)=M_g(t)/M_h(t)$ 
for $t < t_{\rm sat}$ is consistent with cosmological simulations of \cite{okamoto}.
We have checked that these results, especially $t_{\rm sat}$ and $M_h(t_{\rm sat})$,
are not sensitive to different choices of $\alpha=1$ and 2. 

Using the results for the baseline case with $t_{\rm sat}\approx 4.8$~Gyr
and the present orbital parameters of Fornax,
we have related $\Delta F(t)$ at $t>4.8$~Gyr to its orbital motion as an MW satellite
and estimated the effects of ram pressure and tidal interaction with the MW.
The Fornax halo lost $\sim 1/3$ of its CDM through tidal interaction
but reaccreted some of the lost CDM near the end of its first orbital period. This reaccretion 
was responsible for the last episode of significant net gas inflow. Otherwise, gas was 
removed from the star-forming disk as hot gas created by SN heating escaped and was 
then lost from Fornax through ram-pressure stripping and tidal interaction. This lasted 
a few orbital periods and eventually terminated star formation in Fornax.

Our assumed global SFL most likely does not apply to the epoch near
the onset of star formation and prior to reionization. 
By using the median halo growth history $M_h(t)$ prior to the time 
$t_{\rm reion}$ for reionization and assuming the corresponding 
gas mass fraction $f_g(t)=M_g(t)/M_h(t)$ to be the cosmic mean mass 
fraction of baryons $\langle f_b\rangle$, we have estimated 
$t_{\rm reion}$ as the time at which the $f_g(t)$ derived from the assumed SFL 
becomes equal to $\langle f_b\rangle$. We have obtained 
$t_{\rm reion}\sim 0.52$ (0.49)~Gyr corresponding to
$z_{\rm reion}\sim 9.4$ (9.8) for the baseline case with 
$t_{\rm sat}\approx 4.8$ (7.8)~Gyr,
$t_{\rm reion}\sim 0.88$~Gyr ($z_{\rm reion}\sim 6.3$) for the case of $\alpha=1$
and $t_{\rm reion}\sim 0.36$~Gyr ($z_{\rm reion}\sim 12.3$) for the case of $\alpha=2$.
These estimates of $z_{\rm reion}$ are close to $z=8^{+3}_{-2}$ found for reionization
of the IGM surrounding MW dwarf galaxies in simulations by \cite{busha}.
Our estimate of the reionization time can be improved significantly by using detailed 
CDM simulations to estimate halo evolution instead of the model of \cite{zhao}, which
only gives the median halo growth history. We note that data on SFH are available for 
other dwarf galaxies of the Local Group (e.g., \citealt{deboer2,deboer3,weisz}). 
Combining CDM simulations with our approach to study these systems may provide an 
interesting probe of reionization.

In conclusion, we have provided a reasonable picture for the halo evolution and global 
gas dynamics of Fornax based on empirical data. The input to our model includes the 
SFH, the mass enclosed within the half-light radius, and the present-day orbit of Fornax.
Essential to our approach is the SFH and we have given only an approximate
treatment of the correlated errors in each time bin to obtain a smooth SFH.
We emphasize that a rigorous treatment should be carried out to make our approach
more quantitative. Our approach can be extended to other dSphs for which the data 
listed above are available, e.g., Sculptor and Carina.
Our results on $t_{\rm sat}$ and $M_h(t_{\rm sat})$ may help identify the halos that are
most likely to host the observed dSphs in CDM simulations of the formation of 
an MW-like galaxy. Our results on $M_g(t)$ and $\Delta F(t)$ may be used along with
the data on SFHs to develop models of chemical evolution for dSphs 
(e.g., \citealt{qw} and references therein). 
Our empirical approach also serves to illustrate how different sets of observational data 
may be integrated to provide a coherent description of the evolution of individual dSphs. 
We hope that this would help motivate dedicated cosmological simulations with gas 
physics to reproduce the observed dSphs with detailed star formation and chemical 
evolution histories, as well as present-day mass distributions and orbits. Such 
simulations are the best way to resolve the TBTF problem.

\section*{Acknowledgments}
We thank Thomas de Boer for providing the data on Fornax's star formation history
and for advice on treating uncertainties in the data. We also thank an anonymous 
referee for constructive criticisms that helped improve our paper.
We have benefited from discussions with Fabio Governato, Evan Skillman, and 
Jerry Wasserburg. We are grateful to Conrad Chan and Alexander Heger for help 
with the quadratic-spline fit. Z.Y. and Y.Z.Q. acknowledge the hospitality of 
the Institute of Nuclear and Particle Physics at Shanghai Jiao Tong University 
where part of the work was done. This work was supported in part by 
the US DOE (DE-FG02-87ER40328), the NSFC (11320101002), 
973-Project 2015CB857000, Shanghai Key Laboratory Grant No. 11DZ2260700, and
the CAS/SAFEA International Partnership Program for Creative Research Teams 
(KJCX2-YW-T23).

\appendix
\section{Fitting Fornax's SFR to Data}
\label{sec-fit}
The data of \cite{deboer} on Fornax's SFH were binned according to the age $t_*$ 
of stars. We label the bin boundaries as $t_0<t_1<\cdots<t_n$ in terms of 
the time of star formation $t=t_u-t_*$, where $t_u$ is the age of the universe and
$t=0$ corresponds to the big bang. We obtain a smooth SFR $\psi(t)$ from a 
quadratic-spline fit to the data that conserves the total number of stars formed in 
each bin and guarantees the continuity of $\psi(t)$ and $\psi'(t)\equiv d\psi/dt$
(see e.g., C. Chan et al., in preparation).
Specifically, for the $i$th bin $[t_{i-1},t_i]$ with an average SFR $\bar\psi_i$ given
by the data, we seek 
\begin{equation}
\psi(t)=\psi(t_{i-1})+p_i(t-t_{i-1})+q_i(t-t_{i-1})^2,
\label{eq-fit}
\end{equation}
where $p_i$ and $q_i$ are coefficients to be determined. The conservation of the 
total number of stars formed in this bin requires
\begin{equation}
\psi(t_{i-1})=\bar\psi_i-\frac{1}{2}p_i\Delta_i-\frac{1}{3}q_i\Delta_i^2,
\label{eq-area}
\end{equation}
where $\Delta_i\equiv t_i-t_{i-1}$. In addition, Eq.~(\ref{eq-fit}) gives
\begin{eqnarray}
\psi'(t_{i-1})&=&p_i,\label{eq-pi}\\
\psi'(t_i)&=&p_i+2q_i\Delta_i.\label{eq-qi}
\end{eqnarray}
Inspection of Eqs.~(\ref{eq-fit})--(\ref{eq-qi}) shows that $\psi(t)$ for all $t$ 
in the $i$th bin can be obtained once $\psi'(t_{i-1})$ and $\psi'(t_i)$ are
determined.

Equation~(\ref{eq-fit}) gives
\begin{equation}
\psi(t_i)=\psi(t_{i-1})+p_i\Delta_i+q_i\Delta_i^2.\label{eq-pti}
\end{equation}
Substituting $\psi(t_{i-1})$ from Eq.~(\ref{eq-area}) into the above equation, 
we obtain
\begin{equation}
\psi(t_i)=\bar\psi_i+\frac{1}{2}p_i\Delta_i+\frac{2}{3}q_i\Delta_i^2.
\label{eq-pti2}
\end{equation}
All of the equations for the $i$th bin can be generalized to other bins. For example, 
Eq.~(\ref{eq-area}) can be rewritten as
\begin{equation}
\psi(t_i)=\bar\psi_{i+1}-\frac{1}{2}p_{i+1}\Delta_{i+1}-\frac{1}{3}q_{i+1}\Delta_{i+1}^2.
\label{eq-pti3}
\end{equation}
Combining Eqs.~(\ref{eq-pti2}) and (\ref{eq-pti3}) gives
\begin{equation}
\frac{1}{2}p_i\Delta_i+\frac{2}{3}q_i\Delta_i^2+\frac{1}{2}p_{i+1}\Delta_{i+1}
+\frac{1}{3}q_{i+1}\Delta_{i+1}^2=\bar\psi_{i+1}-\bar\psi_i.
\end{equation}
Using Eqs.~(\ref{eq-pi}) and (\ref{eq-qi}) to eliminate $p_i$, $q_i$, $p_{i+1}$, and 
$q_{i+1}$ from the above equation, we obtain
\begin{equation}
\frac{\Delta_i}{6}\psi'(t_{i-1})+\frac{\Delta_i+\Delta_{i+1}}{3}\psi'(t_i)
+\frac{\Delta_{i+1}}{6}\psi'(t_{i+1})=\bar\psi_{i+1}-\bar\psi_i,
\end{equation}
which can be rewritten as a matrix equation for $\{\psi'(t_i),0\leq i\leq n\}$. 
The unknown $\{\psi'(t_i),0<i<n\}$ can be solved from this equation by specifying
$\psi'(t_0)$ and $\psi'(t_n)$.
 
We take $\psi'(t_0)=\psi'(t_n)=0$ to obtain the smooth $\psi(t)$ that is shown in 
Fig.~\ref{fig-sfr} along with the data. We use $t_u=13.79$~Gyr for the adopted
cosmology to convert the stellar age $t_*$ into the time $t$ of star formation.
The youngest stars covered by the data have $t_*=0.25$--0.5~Gyr 
corresponding to $t=13.29$--13.54~Gyr. Considering that there is no current star 
formation in Fornax, we have included $13.54\leq t\leq 13.79$~Gyr in the fit as 
the $n$th bin with $\bar\psi_n=0$. To ensure that no net stars were formed 
in this bin, the corresponding $\psi(t)$ must drop from positive to negative values.
We take the part of the fit with $\psi(t)\geq 0$ and ignore the unphysical part with 
$\psi(t)<0$. This gives an estimate of $t\approx 13.6$~Gyr for the end of star 
formation in Fornax. The oldest stars covered by the data have $t_*=13$--14~Gyr. 
So formally our first bin corresponds to $-0.21\leq t\leq 0.79$~Gyr. However, 
the stellar ages in this bin have uncertainties of $\sim 2.5$~Gyr \citep{deboer}. 
Therefore, we take the centroid of this bin at $t=0.29$~Gyr as 
approximately the onset of star formation in Fornax and
ignore the part of the fit for $t<0.29$~Gyr. The uncertainties associated with
the first and last bins for our fit represent end effects that are mostly restricted
to the corresponding periods. Our main results are based on the fitted $\psi(t)$
outside these periods.

\section{Determination of Fornax's Halo Mass}
\label{sec-mh}
In discussing halo evolution, it is convenient to use redshift $z$. 
For a flat $\Lambda$CDM cosmology, we have
\begin{equation}
\frac{dz}{dt}=-(1+z)\sqrt{\frac{8\pi G\rho_{\rm cr}(z)}{3}},
\end{equation}
where
\begin{equation}
\rho_{\rm cr}(z)\equiv\rho_{\rm cr}(0)\left[\Omega_m(1+z)^3+\Omega_\Lambda\right]
\end{equation}
is the critical density at redshift $z$, $\rho_{\rm cr}(0)\equiv3H_0^2/(8\pi G)$
is the critical density at the present time, $H_0=100h$~km~s$^{-1}$~Mpc$^{-1}$ 
is the Hubble parameter, and $\Omega_m=\Omega_{\rm CDM}+\Omega_b$ and 
$\Omega_\Lambda$ are the fractional contributions to $\rho_{\rm cr}(0)$ from 
non-relativistic matter (CDM plus baryons) and the cosmological constant, 
respectively. Throughout this paper, we adopt $h=0.69$, $\Omega_m=0.29$
($\Omega_{\rm CDM}=0.243$, $\Omega_b=0.047$), and 
$\Omega_\Lambda=0.71$. In calculating the growth of a halo, 
we use a primordial power spectrum with a spectral index $n_s=0.96$ and the 
transfer function of \cite{eh98} with a present temperature of 2.726~K for the 
cosmic microwave background to obtain the linear power spectrum, the amplitude
of which is fixed by $\sigma_8=0.82$. The above cosmological parameters are 
consistent with the final analysis of the WMAP experiment \citep{wmap}. 

In addition to the total mass of a halo, another important parameter is its virial radius. 
For a halo of mass $M_h$ collapsing at redshift $z$, its virial radius $r_{\rm vir}$
\citep{bryan} is defined through
\begin{equation}
M_h=\frac{4\pi}{3}r_{\rm vir}^3\rho_{\rm cr}(z)\Delta_c,
\label{eq-rvir}
\end{equation}  
where
\begin{equation}
\Delta_c=[18\pi^2+82(\Omega_z-1)-39(\Omega_z-1)^2],
\end{equation}
and
\begin{equation}
\Omega_z\equiv\frac{\Omega_m(1+z)^3}{\Omega_m(1+z)^3+\Omega_\Lambda}.
\end{equation}
The mass distribution of a halo can be approximated by the Navarro-Frenk-White 
(NFW) density profile
\begin{equation}
\rho(r)=\frac{\rho_0}{(r/r_s)[1+(r/r_s)]^2},
\label{eq-nfw}
\end{equation}
where $\rho_0$ and $r_s$ are two parameters characteristic of the halo.
Using the above profile, we obtain
\begin{equation}
M_h\equiv\int_0^{r_{\rm vir}}4\pi r^2\rho(r)dr=
4\pi\rho_0r_s^3\left[\ln(1+c)+\frac{c}{1+c}\right],
\label{eq-mhnfw}
\end{equation}
where $c\equiv r_{\rm vir}/r_s$ is referred to as the concentration parameter.
For a halo of mass $M_h$ collapsing at redshift $z$, the concentration parameter 
$c(M_h,z)$ can be estimated from the model of \cite{zhao}.

We assume that the density profile of the Fornax halo was approximately fixed
at $t_{\rm sat}\approx 4.8$~Gyr ($z_{\rm sat}\approx 1.33$) when the halo 
mass reached $M_h(z_{\rm sat})$. As discussed in \S\ref{sec-sat}, tidal interaction
with the MW would not have affected the density profile within the half-light radius 
$r_{1/2}$ for the present-day Fornax. Using the NFW profile and 
Eq.~(\ref{eq-mhnfw}), we obtain the mass $M(<r_{1/2})$ enclosed within 
$r_{1/2}$ as
\begin{equation}
M(<r_{1/2})=M_h(z_{\rm sat})\frac{f(cr_{1/2}/r_{\rm vir})}{f(c)},
\label{eq-mr12}
\end{equation}
where
\begin{equation}
f(x)=\ln(1+x)+\frac{x}{1+x},
\end{equation}
and both $c$ and $r_{\rm vir}$ are evaluated for $M_h(z_{\rm sat})$.
Estimating $c(M_h,z_{\rm sat})$ with the model of \cite{zhao}, we 
solve Eq.~(\ref{eq-mr12}) by iteration to obtain
$M_h(z_{\rm sat})\approx 1.8\times 10^9\,M_\odot$ for
$M(<r_{1/2})=7.39^{+0.41}_{-0.36}\times 10^7\,M_\odot$ and
$r_{1/2}=944\pm53$~pc \citep{wolf}.

\section{Characteristic Parameters of a Halo}
\label{sec-vir}
We define some useful quantities (e.g., \citealt{barkana}) for discussing 
the evolution of a halo. For a halo of total mass $M_h$ collapsing at
redshift $z$, its virial radius $r_{\rm vir}$ is defined by Eq.~(\ref{eq-rvir})
and can be evaluated as
\begin{equation}
r_{\rm vir}=\frac{7.85}{1+z}\left(\frac{M_h}{10^8\,M_\odot}\right)^{1/3}
\left(\frac{18\pi^2}{\Delta_c}\right)^{1/3}
\left(\frac{\Omega_z}{\Omega_mh^2}\right)^{1/3}\ {\rm kpc}.
\end{equation}
Its circular velocity is
\begin{equation}
v_{\rm circ}=\sqrt{\frac{GM_h}{r_{\rm vir}}}
=20.7\left(\frac{M_h}{10^8\,M_\odot}\right)^{1/2}
\left(\frac{\rm kpc}{r_{\rm vir}}\right)^{1/2}\ {\rm km}\ {\rm s}^{-1},
\label{eq-vcirc}
\end{equation}
and its virial temperature is
\begin{eqnarray}
T_{\rm vir}&=&\frac{\mu m_pv_{\rm circ}^2}{2k}\nonumber\\
&=&3.32\times10^3(1+z)\mu\left(\frac{M_h}{10^8\,M_\odot}\right)^{2/3}
\left(\frac{\Delta_c}{18\pi^2}\right)^{1/3}\nonumber\\
&&\times\left(\frac{\Omega_mh^2}{\Omega_z}\right)^{1/3}\ {\rm K},
\label{eq-tvir}
\end{eqnarray}
where $\mu$ is the mean molecular weight for those electrons, nuclei,
and atoms that contribute to the gas pressure, and $k$ is the Boltzmann constant.
We take the primordial mass fractions of protons and $^4$He nuclei to be 0.75 and 
0.25, respectively, and use $\mu=1.23$ and 0.59 for neutral and fully ionized gas, 
respectively. Following \cite{busha}, we take $\mu=1.23$ for 
$T_{\rm vir}<1.5\times 10^4$~K and $\mu=0.59$ for 
$T_{\rm vir}>1.5\times 10^4$~K. The transition between these two regimes is
assumed to occur at a fixed $T_{\rm vir}=1.5\times 10^4$~K.
Using $M_h(t)$ for the Fornax halo shown in
Fig.~\ref{fig-mh}, we show the corresponding $r_{\rm vir}(t)$ and $T_{\rm vir}(t)$
in Fig.~\ref{fig-vir}.

\label{lastpage}
\end{document}